\documentclass[nofootinbib,superscriptaddress,a4paper,twocolumn, floatfix]{revtex4-2}
\usepackage[english]{babel}
\usepackage[utf8]{inputenc}

\usepackage{amsthm}
\usepackage{mathtools}
\usepackage{physics}
\usepackage{xcolor}
\usepackage{graphicx}
\usepackage[left=23mm,right=13mm,top=35mm,columnsep=15pt]{geometry} 
\usepackage{adjustbox}
\usepackage{placeins}
\usepackage[T1]{fontenc}
\usepackage{lipsum}
\usepackage{csquotes}

\usepackage[pdftex, pdftitle={Article}, pdfauthor={Author}]{hyperref} 
\usepackage{xcolor}
\hypersetup{
    colorlinks,
    linkcolor={red!50!black},
    citecolor={blue!50!black},
    urlcolor={blue!80!black}
}
\usepackage{booktabs}
\usepackage{multirow}
\usepackage{listings}
\usepackage{xcolor}

\definecolor{codegreen}{rgb}{0,0.6,0}
\definecolor{codegray}{rgb}{0.5,0.5,0.5}
\definecolor{codepurple}{rgb}{0.58,0,0.82}
\definecolor{tqblue}{HTML}{08293d}
\definecolor{backcolour}{HTML}{fefdf5}

\lstdefinestyle{mystyle}{
    backgroundcolor=\color{backcolour},   
    commentstyle=\color{codegreen},
    keywordstyle=\color{magenta},
    numberstyle=\tiny\color{codegray},
    stringstyle=\color{codepurple},
    basicstyle=\ttfamily\footnotesize\color{tqblue},
    breakatwhitespace=false,         
    breaklines=true,
    postbreak=\mbox{\textcolor{magenta}{$\hookrightarrow$}\space},                 
    captionpos=b,                    
    keepspaces=true,                 
    numbers=left,                    
    numbersep=5pt,                  
    showspaces=false,                
    showstringspaces=false,
    showtabs=false,                  
    tabsize=2
}

\newcommand{\expvals}[2]{\ensuremath{\langle #1 \rangle_{#2}}}
\begin{document}

\title{A Feasible Approach for Automatically Differentiable Unitary Coupled-Cluster on Quantum Computers}

\author{Jakob~S.~Kottmann}\textbf{}
\email[E-mail:]{jakob.kottmann@utoronto.ca}
\affiliation{{Chemical Physics Theory Group, Department of Chemistry, University of Toronto, Canada.}}
\affiliation{Department of Computer Science, University of Toronto, Canada.}

\author{Abhinav Anand}\textbf{}
\affiliation{{Chemical Physics Theory Group, Department of Chemistry, University of Toronto, Canada.}}

\author{Alán Aspuru-Guzik}
\email[E-mail:]{aspuru@utoronto.ca}
\affiliation{{Chemical Physics Theory Group, Department of Chemistry, University of Toronto, Canada.}}
\affiliation{Department of Computer Science, University of Toronto, Canada.}
\affiliation{Vector Institute for Artificial Intelligence, Toronto, Canada.}
\affiliation{Canadian  Institute  for  Advanced  Research  (CIFAR)  Lebovic  Fellow,  Toronto,  Canada}

\date{\today} 

\begin{abstract}
We develop computationally affordable and encoding independent gradient evaluation procedures for unitary coupled-cluster type operators, applicable on quantum computers.
We show that, within our framework, the gradient of an expectation value with respect to a parameterized $n$-fold fermionic excitation can be evaluated by four expectation values of similar form and size, whereas most standard approaches based on the direct application of the \textit{parameter-shift-rule} come with an associated cost of $\mathcal{O}\left(2^{2n}\right)$ expectation values.
For real wavefunctions, this cost can be further reduced to two expectation values.
Our strategies are implemented within the open-source package \textsc{tequila} and allow blackboard style construction of differentiable objective functions.
We illustrate initial applications for electronic ground and excited states.
\end{abstract}

\maketitle

\section{Introduction}

The proposition of using the quantum phase estimation algorithm to extract eigenenergies of electronic Hamiltonians with the help of quantum computers\cite{aspuru2005simulated} resulted in various new research ideas for quantum chemistry on quantum computers. One such direction was the introduction and successful demonstration of the quantum variational eigensolver (VQE)~\cite{peruzzo2014variational, McClean2016theoryofvqe}, a new method for approximating eigenenergies with current and near-term quantum hardware in mind.
Variational quantum algorithms apply the variational principle to expectation values of a parameterized quantum circuit $U\left(\theta\right)$ and a qubit Hamiltonian $H$ by optimizing the parameters with a classical optimization algorithm
\begin{align}
    \min_{\theta} \left(\langle H \rangle_{U(\theta)}\right) \equiv \min_\theta\left( \langle 0 \rvert U^\dagger\left(\theta\right) H  U\left(\theta\right) \lvert 0\rangle\right).
\end{align}
The original proposal of VQE inspired development of numerous new variational quantum algorithms for estimating energies of quantum chemical and many-body models on quantum computers (see Refs.~\cite{cao2019quantumreview, mcardle2020quantumreview} for recent reviews) as well as in various other fields including quantum machine learning~\cite{schuld2020effect,biamonte2017quantum, PerezSalinas2020datareuploading, schuld2019quantum, schuld2020classifiers, romero2017quantum, anand2020experimental, kristensen2019artificial}, combinatorial optimization~\cite{farhi2014quantum} and quantum optics~\cite{kottmann2020quantum, VQU, sawaya2020resource}.
These developments in combination with the recent availability of
open access quantum computers~\cite{ibmq}, and significant improvements of currently available quantum hardware~\cite{arute2019quantum, wang201818, rudolph2017optimistic} are currently clearing the path towards
Feynman's original idea of simulating physics with quantum computers~\cite{feynman1982simulating} leveraging this powerful tool to elucidate challenging chemical processes~\cite{reiher2017elucidating}. \\

The use of gradient based optimization method is not always the canonical choice for variational quantum algorithms. 
Its applicability depends strongly on the number of used parameters and the form of the parameterized gates.
For specific parameterized gates the evaluation of analytical gradients of expectation values becomes comparably cheap by applying the \textit{parameter-shift-rule}~\cite{schuld2019evaluating} where the expectation value has to be evaluated two times with shifted parameters leading to elegant and computationally feasible implementations of automatic differentiation within quantum algorithms pioneered within \textsc{pennylane}~\cite{bergholm2018pennylane}.\\

In this work, we extend the framework of automatically differentiable quantum algorithms to unitary coupled-cluster in its separated framework~\cite{evangelista2019, izmaylov2020order, grimsley2019adaptive}.
We develop procedures to calculate the gradients of expectation values generated by arbitrary $n$-fold fermionic excitation operators by compiling it into linear combinations of expectation values that can be evaluated on quantum computers. Similar to the \textit{parameter-shift-rule}, those expectation values include the original excitation unitaries with shifted parameters but require an additional unitary generated by the nullspace projector of the corresponding excitation generator. If all underlying wavefunctions are real we show that two of the four expectation values become equivalent, which reduces the computational cost to two. The main results are summarized in Tab.~\ref{tab:generator_overview}.
All developed techniques are implemented in the open-source package \textsc{tequila}\cite{tequila} and allow blackboard style construction of automatically differentiable objectives constructed from fermionic excitation operators and other quantum gates. We present explicit code examples and initial application for electronic ground and excited states.\\

\begin{table}[htbp]
    \centering
    \begin{tabular}{ccc}
    \toprule
    Generator Form  & Gradient Cost & Strategy \\
    \midrule
        \multirow{2}{*}{$\displaystyle G_{\mathbf{p}\mathbf{q}} =\sum_{\mathbf{i}} c_{\mathbf{i}}\sigma_{\mathbf{i}}$ }
        & \multirow{2}{*}{$\displaystyle\mathcal{O}\left(2^{2n}\right)$} & shift-rule \\
        & & Eq.~\eqref{eq:shift_rule}  \\ 
    \midrule
        \multirow{2}{*}{$\displaystyle
        G_{\mathbf{p}\mathbf{q}}=\frac{1}{2}\left(G_+ + G_-\right)$ }
        & \multirow{2}{*}{4} 
        & fermionic-shift\\
        & & Eq.~\eqref{eq:gradient_decompositon_highest_level}  \\
    \midrule
    \multicolumn{3}{c}{Real Wavefunctions}\\
    \midrule
        \multirow{2}{*}{$\displaystyle G_{\mathbf{p}\mathbf{q}}=\frac{1}{2}\left(G_+ + G_-\right)$}
        &\multirow{2}{*}{2} & fermionic-shift \\
        & & Eq.~\eqref{eq:gradient_decompositon_real_wfn}\\
    \midrule
    \multicolumn{3}{c}{Generator Approximation}\\
    \midrule
        \multirow{2}{*}{$\displaystyle G_{\mathbf{p}\mathbf{q}}\approx G_\pm$ }  
        & \multirow{2}{*}{2} 
        & shift-rule \\
        & & Eq.~\eqref{eq:shift_rule}\\
    \bottomrule
    \end{tabular}
    \caption{Overview over the used fermionic generators in this work. The gradient cost is given as a factor with respect to the cost of the evaluation of the original expectation value. See Eqs.~\eqref{eq:Gn},~\eqref{eq:Gpm_definition} and~\eqref{eq:P0_fermionic} for the definition of the $G$ and $G\pm$ generators and how to construct them using the nullspace projector $P_0$.}
    \label{tab:generator_overview}
\end{table}

\section{Methodology}\label{sec:Method}
The framework of unitary coupled-cluster allows the construction of quantum circuits as a product of unitaries that create fermionic excitations within the wavefunction.
Those unitary operations can be described in terms of their hermitian generators $G$ as
\begin{align}
    U(\theta) = e^{-i\frac{\theta}{2} G}
\end{align}
where the generators for singles, doubles and $n$-fold fermionic excitations are given by
\begin{align}
    G_{pq} &= i(a_p^\dagger a_q - a^\dagger_q a_p)\label{eq:G1}\\
    G_{pqrs} &= i(a_p^\dagger a_q a_r^\dagger a_s - h.c.)\label{eq:G2} \\
    G_{\mathbf{p}\mathbf{q}} &= i( \prod_{i=1}^n a_{p_i}^\dagger a_{q_i} - h.c.)\label{eq:Gn},   
\end{align}
with $a^\dagger,a$ being the usual anti-commuting fermionic creation and annihilation operators~\cite{cao2019quantumreview,helgaker2014molecular, shavitt2009many, surjan2012second, jorgensen2012second}.
Given the condition that the generator $G$ of a parametrized unitary $U(\theta)=e^{-i\theta G}$ has only two distinct eigenvalues $\pm r$, Schuld \textit{et.al.}~\cite{schuld2019evaluating} showed that the direct measurement of the \textit{analytical} gradient of expectation values $\bra{\Psi}\hat{O}\ket{\Psi} \equiv \expvals{O}{U_\Psi}$, formed from unitaries $U_\Psi$ including $U(\theta)$, and some hermitian operator $\hat{H}$, can be achieved as
\begin{align}
    \frac{\partial \expvals{H}{XU(\theta)Y}}{\partial \theta} = r \left( \expvals{H}{XU(\theta + s)Y} - \expvals{H}{XU(\theta -s)Y} \right)\label{eq:shift_rule}
\end{align}
with $s = \frac{\pi}{4r}$ and where we used $U_\Psi(\theta) = XU(\theta)Y$ to illustrate the differentiable gate $U(\theta)$ within a larger abstracted unitary symbolized by $X$ and $Y$.
Further improvements on this gradient evaluation technique in the context of stochastic sampling~\cite{banchi2020measuring}, higher order derivatives~\cite{mari2020estimating}, classical simulation~\cite{jones2020efficient} and noisy evaluations~\cite{meyer2020variational} have been developed, making it a standard tool within variational quantum algorithms.
If the condition of having only two distinct eigenvalues does not hold for the generator, the analytical gradient can still be obtained by more sophisticated techniques. One is the decomposition of $U(\theta)$ into a product of directly differentiable unitaries~\cite{crooks2019gradients} and another involves the execution of the unitary controlled by an additional ancillary qubit~\cite{schuld2019evaluating, romero2018strategies}.
The fermionic generators of Eqs~\eqref{eq:G1},~\eqref{eq:G2} and~\eqref{eq:Gn} do not fulfill the necessary condition to be directly differentiable but their gradients can still be evaluated by these extended approaches where the latter has been demonstrated by Romero \textit{et.al}~\cite{romero2018strategies}. The decomposition approach can be realized in a straightforward way described in the following.
In order to be executable on a general quantum computer, 
the $n$-fold excitation generators of Eq.~\eqref{eq:Gn} are transformed into qubit operators by writing them as a linear combination of tensor-products of Pauli matrices $\boldsymbol{\sigma}_i$, often referred to as Pauli strings
\begin{align}
    G_{\mathbf{p}\mathbf{q}} = \sum_i c_{i(\mathbf{p}, \mathbf{q})} \boldsymbol{\sigma}_{i(\mathbf{p}, \mathbf{q})}
\end{align}
where the length of the sum and the individual form of the coefficients $c$ and Pauli strings $\boldsymbol{\sigma}$ depends on the generator and the chosen transformation.
The Pauli strings arising from an individual generator commute amongst each other~\cite{romero2018strategies} so that the qubit unitary generated from those Pauli strings decomposes into a product of multi-Pauli rotations $e^{i\frac{\theta}{2}G} \rightarrow \prod_k e^{i\frac{\theta}{2} c_k\boldsymbol{\sigma}_k}$. Since individual Pauli strings $\boldsymbol{\sigma}$ are self-inverse they only have two distinct eigenvalues ($\pm 1$) making the unitaries generated by them directly differentiable.
The gradient can then be evaluated by combining the product rule of calculus with the \textit{parameter-shift-rule} of Eq.~\eqref{eq:shift_rule}.
Usually the generators of $n$-fold fermionic excitation operators are transformed into $\mathfrak{O}(2^{2n-1})$ Pauli strings by the Jordan-Wigner or Bravyi-Kitaev transformation.~\cite{bravyi2002, seeley2012, tranter2015, tranter2018} This will lead to the evaluation of $\mathfrak{O}\left(2^{2n}\right)$ expectation values in order to obtain their gradients by decomposition into directly differentiable gates.\\

\begin{figure}[htbp]
    \centering
    \includegraphics[width=0.4\textwidth]{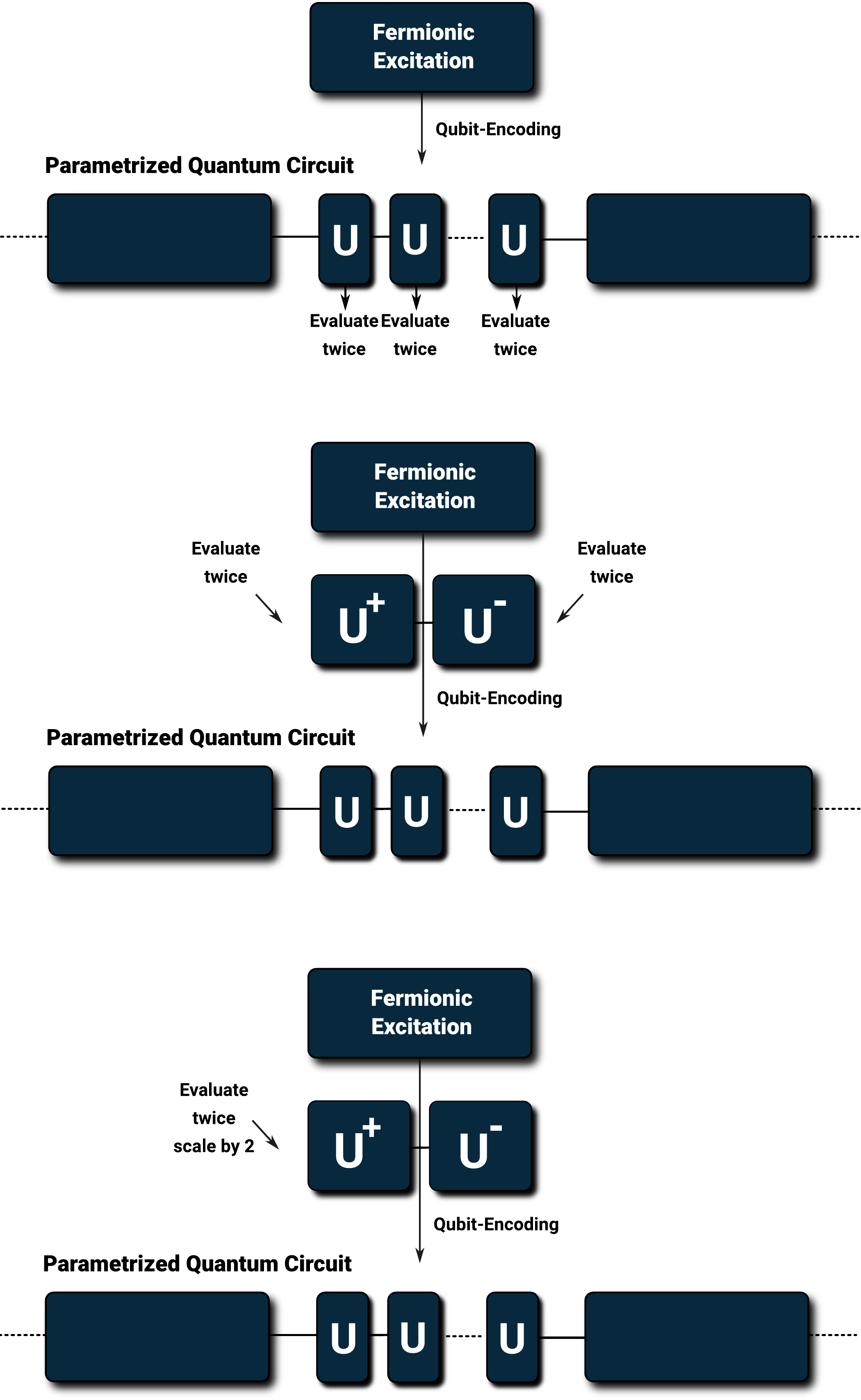}
    \caption{\textbf{Automatic differentiation for fermionic operators:} Schematic overview over standard approaches acting on the qubit level using Eq.~\eqref{eq:shift_rule} (top) and the general gradient evaluation schemes on the fermionic level according to Eq.~\eqref{eq:gradient_decompositon_highest_level} (middle) as well as for real wavefunctions according to Eq.~\eqref{eq:gradient_decompositon_real_wfn} (bottom). See also Tab.~\ref{tab:generator_overview} for a general overview and Eq.~\eqref{eq:fermionic_shift_gate} for the definition of the fermionic shift gates $U^{\alpha}_\pm$.}
    \label{fig:algorithm_overview}
\end{figure}

In the following we will show how to evaluate the gradients of $n$-fold fermionic excitation operators with a constant cost factor of 4 instead of $\mathcal{O}\left(2^{2n}\right)$. For real wavefunctions this cost factor can be lowered to 2, making it equivalent in cost as the simplest finite difference stencils. We will start with generalized operators - that don't have to be fermionic excitations - and afterwards show how those operators can be constructed in the fermionic representation. For the convenience of the reader we summarize our findings in Tab.~\ref{tab:generator_overview} and illustrate the implementation of the automatic differentiation within \textsc{tequila}~\cite{tequila} in Fig.~\ref{fig:algorithm_overview}.
\\

\subsection{Generator Decomposition}
The generators $G_{\mathbf{p}\mathbf{q}}$ ~\eqref{eq:Gn} of fermionic excitations act only non-trivially on states with all $\mathbf{p}$ orbitals empty and all $\mathbf{q}$ orbitals occupied or vice versa. On all other states the generators act as zeroes, so that the corresponding unitary acts as identity operator. As a consequence, the generators have three distinct eigenvalues, $\pm1$ and 0 (see the appendix for more details).
We can formally write the generator as the sum over the projectors $P_{+}, P_{-}$ and $P_0$ that project onto the spaces spanned by the eigenfunctions of $G$ multiplied by their corresponding eigenvalues $\pm1$ and 0. The generator can then formally be written as
\begin{align}
    G = P_{+} - P_{-}.
\end{align}
Note that the projector $P_0$ is not included due to its zero eigenvalue.
The individual projectors $P_{\pm}$ themselves have two distinct eigenvalues ($0$ and $1$) and commute amongst themselves due to the orthogonality of the eigenstates. The generated unitary can then be split into two directly differentiable parts
\begin{align}
    e^{-i\frac{\theta}{2}G} = e^{-i\frac{\theta}{2}P_+}e^{+i\frac{\theta}{2}P_-}
\end{align}
reducing the gradient cost to a constant factor of 4 when the \textit{parameter-shift-rule} is combined with the product rule.
In Fig.~\ref{fig:algorithm_overview} this procedure is schematically illustrated and compared with the standard approach that applies the \textit{parameter-shift-rule} on the qubit level to each Pauli string individually.
An alternative way to split the generator is by adding (subtracting) the nullspace projector $P_0$ to the generators
\begin{align}
    G = \frac{1}{2}\left(G_+ + G_-\right)\label{eq:split_G_Gpm}
\end{align}
introducing the self-inverse generators
\begin{align}
    &G_\pm = G \pm P_0.\label{eq:Gpm_definition}\\
    &G_\pm^2 = P_+ + P_- + P_0 = 1,
\end{align}
where the last equation holds due to the completeness and orthonormality of the eigenspace projectors.
Both ways of splitting the generator will result in similar quantum circuits since the unitaries generated by $G_\pm$ just differ by a phase from the unitaries generated by $P_\pm$ (see Eq.~\eqref{eq:Ppm_decomposition}) and we will mostly stick with the $G_\pm$ generators since they share many properties with single qubit rotations and can be treated as generalized multi-qubit rotations
\begin{align}
    e^{-i\frac{\theta}{2}\frac{1}{2}G\pm} = \cos\left(\frac{\theta}{4}\right) - i \sin\left(\frac{\theta}{4}\right)G\pm.\label{eq:Gpm_decomposition}
\end{align}
Using this formula we can express how a unitary, generated by a fermionic excitation, acts in a closed analytical form (see the appendix for a detailed derivation)
\begin{align}
    U\left(\theta\right) &= e^{-i\frac{\theta}{2}G} \label{eq:G_decomposition}\\
    &= \cos\left(\frac{\theta}{2}\right)\mathbf{1} - i\sin\left(\frac{\theta}{2}\right)G + \left(1-\cos\left(\frac{\theta}{2}\right)\right)P_0.\nonumber
\end{align}
When acting on a specific electronic configuration $\ket{\Phi}$ the unitary acts as a unit operator if the configuration is in the nullspace of the generator, or, as a rotation between the original configuration and the $n$-fold excited configuration
\begin{align}
    U\left(\theta\right)\ket{\Phi}=\begin{cases}
        \ket{\Phi}, \quad P_0\ket{\Phi} = \ket{\Phi}\\
    \left(\cos\left(\frac{\theta}{2}\right) -i \sin\left(\frac{\theta}{2}\right)G\right)\ket{\Phi}, \quad \text{else}
    \end{cases}.
\end{align}
Note that the generated superpositions are real due to the definition of the hermitian generators G in Eq.~\eqref{eq:Gn}.

\subsection{Exact Analytical Gradients}
With the splitting of the generators introduced in Eq.~\eqref{eq:split_G_Gpm}, analytical gradients of $n$-fold fermionic excitations can be evaluated by combining the parameter-shift-rule with the product rule of calculus. Due to the product rule, this will lead to a gradient cost factor of 4 meaning that 4 expectation values with similar cost to the original expectation value have to be evaluated
\begin{align}
    \frac{\partial \expvals{H}{XU(\theta)Y}}{\partial \theta} &= r \sum_{\alpha \in \left\{+,-\right\}}\left(\expvals{H}{XU^\alpha_+Y} - \expvals{H}{XU^\alpha_-Y}\right)\label{eq:gradient_decompositon_highest_level}
\end{align}
where $U^\alpha_\pm$ denotes the fermionic unitaries with shifted $G_\alpha$ part. Following Ref.~\cite{schuld2019evaluating} the shift will be $s=\frac{\pi}{4r} = \pi$ with $r=\frac{1}{4}$.\\

Using an automatically differentiable framework, as for example offered in \textsc{tequila}~\cite{tequila}, this gradient evaluation procedure can be implemented in a straightforward way.
Using the most straightforward realization of this scheme, the explicit implementation of the unitary that is being differentiated will require approximately twice the number of native quantum gates as in the original expectation value since the two generators $G_\pm$ decompose into a similar number of Pauli strings as the original generator $G$. Note however that this only holds for the unitary that is being differentiated so the overall gate count for the gradients will just grow by a small constant that can be mitigated by applying more advanced compiling and gate fusion techniques.
We will now show one simplification that can already be done on the fermionic level and that will result in the original unitary plus a unitary generated by it's nullspace projector $P_0$.
Consider the first part of the product rule in Eq.~\eqref{eq:gradient_decompositon_highest_level}, where the parameter-shift is performed on the $G_+$ generator, \textit{i.e.} $\alpha=+$. The shifted unitary is then
\begin{align}
    U_\pm^{+} &=e^{-i\frac{1}{4}\left(\theta\pm\pi\right)G_+} e^{-i\frac{\theta}{4}G-}\nonumber\\
    &=e^{-i\frac{1}{2}\left(\theta \pm \frac{\pi}{2}\right)G} e^{-\frac{i}{2}\left(\pm\frac{\pi}{2}\right)P_0},
\end{align}
and if we shift the $G_-$ part, we will arrive at the same expression with inverted sign on the $P_0$ dependent part. In general this fermionic shift operator can be written as a product of the shifted fermionic gate $U_{\pm}=U\left(\theta \pm \frac{\pi}{2}\right)$ and an additional, parameter independent, gate $U_0^\alpha$ generated by the nullspace projector of the generator
\begin{align}
    U_\pm^{\alpha}\left(\theta\right) &= U_\pm\left(\theta\right) U_0^\alpha\label{eq:fermionic_shift_gate}\\
    &=e^{-i\frac{1}{2}\left(\theta \pm \frac{\pi}{2}\right)G} e^{-\alpha\frac{i}{2}\left(\pm\frac{\pi}{2}\right)P_0},\quad \alpha \in \left\{ +1,-1 \right\}.\nonumber 
\end{align}

\subsection{Real Wavefunctions}
In the following we will present a strategy to lower the cost of the exact gradient formula in Eq.~\eqref{eq:gradient_decompositon_highest_level} to only two expectation values under the condition that the involved wavefunctions are real. In particular we demand that for a general quantum circuit $XU(\theta)Y$ the wavefunctions $Y\ket{0}$ are real and that the part denoted by $X$ only generates real superpositions. For pure unitary coupled-cluster type circuits this reduces to the requirement of a real reference wavefunction.
Under this conditions, two parts of the sum in Eq.~\eqref{eq:gradient_decompositon_highest_level} become equivalent and the gradient can be evaluated as
\begin{align}
    \frac{\partial \expvals{H}{XUY}}{\partial \theta} = \frac{1}{2} \left( \expvals{H}{XU^\alpha_+Y} - \expvals{H}{XU^\alpha_-Y} \right)\label{eq:gradient_decompositon_real_wfn}
\end{align}
where $\alpha$ can be freely chosen to be either $+$ or $-$. Note that the shift in the shifted fermionic gate $U_\pm$ is $s=\frac{\pi}{2}$ making the evaluation scheme of Eq.~\eqref{eq:gradient_decompositon_real_wfn} similar to the \textit{parameter-shift-rule} for single qubit rotations (Eq.~\eqref{eq:shift_rule} with $r=\frac{1}{2}$) with the only difference being the $U_0^\alpha$ gate after the shifted unitaries.\\

Lets start with the exact expression~\eqref{eq:gradient_decompositon_highest_level} for the analytical gradient derived in the last section.
We can formally decompose the individual expectation values by inserting the identity as $1 = P_0 + (1-P_0)$ and using the properties
\begin{align}
     &XU^{\alpha}_{\pm}\left(1-P_0\right)Y =XU_{\pm}Y\ket{0}\\
     &XU^\alpha_{\pm}P_0Y = e^{- \alpha i\frac{1}{2}\frac{\pi}{2}}XP_0Y\ket{0}
\end{align}
where we used the idempotency $P_0^2 =P_0$ of the projector and $GP_0 = 0$ resulting from $P_0$ being the nullspace projector of $G$. As in Eq.~\eqref{eq:fermionic_shift_gate}, $U_\pm=e^{-i\frac{1}{2}\left(\theta\pm\frac{\pi}{2}\right)G}$ denotes the shifted fermionic unitary.
The analytical gradient becomes then
\begin{align}
\frac{\partial \expvals{H}{XUY}}{\partial \theta} =& \frac{1}{4}\sum_{\alpha\in\left\{+,-\right\}} \left( \expvals{H}{XU_+Y} - \expvals{H}{XU_-Y} \right. \nonumber\\ 
&+ \left. \left( e^{-\alpha i\frac{\pi}{4}}\langle{0}\rvert Y^\dagger U_+^\dagger X^\dagger H XP_0Y\lvert0 \rangle + h.c \right) \right. \nonumber\\ &- \left. \vphantom{\frac{1}{4}}
\left( e^{\alpha i\frac{\pi}{4}}\langle{0}\rvert Y^\dagger U_-^\dagger X^\dagger H XP_0Y\lvert0 \rangle + h.c \right) \right) \nonumber\\
=& \frac{1}{2}\left( \expvals{H}{XU_+Y} - \expvals{H}{XU_-Y} \right) + R
\end{align}
where the first part is the same formula as for single qubit rotations and the second part denotes the residual $R$ that can be written as
\begin{align}
    &R = \frac{1}{4}\left( c d_+ + c^*d^*_+ - c d_- - c^*d^*_-\right),\nonumber\\
    &\phantom{R}+\frac{1}{4}\left( c^* d_+ + cd^*_+ - c^*d_- - cd^*_-\right)
\end{align}
using $d_\pm= \langle{0}\rvert Y^\dagger U_+^\dagger X^\dagger H XP_0Y\lvert0 \rangle$ and $c=e^{-i\frac{\pi}{4}}$.
Under the assumptions made above, the numbers $d_\pm$ are real numbers resulting in the two terms in the residual $R$ to become identical. Note however, that the residue does not vanish. This means, that for real wavefunctions the two parts of the $\alpha$ sum in Eq.~\eqref{eq:gradient_decompositon_highest_level} become identical hence it is sufficient to evaluate only one part of the product-rule induced sum and scale it by a factor of two leading to the expression in Eq.~\eqref{eq:gradient_decompositon_real_wfn}.

\subsection{Approximations}
If complex wavefunctions are involved, the gradient evaluation scheme of Eq.~\eqref{eq:gradient_decompositon_real_wfn} becomes an approximation. It will however still be exact in cases where the wavefunction $Y\ket{0}$ (using the notation of the previous section) has no overlap with the nullspace of the fermionic generator that is being differentiated.
Other types of approximations could be made for example by approximating the whole generator by either $G_+$ or $G_-$ (or equivalently as $P_\pm$) which will lead to unitaries that are directly differentiable by the original shift rule of Eq.~\eqref{eq:shift_rule}. The generated unitaries will act in the same way as the unitaries generated from the original generators but will introduce phase factors to all nullspace elements of the wavefunction (see Eq.~\eqref{eq:Gpm_decomposition}). If the wavefunction on which the unitaries act are not supported on the nullspace of the original generator $G$, \textit{i.e.} $P_0Y\ket{0} = 0$, the unitaries generated by $G_\pm$ will act identical as unitaries generated by $G$. This leaves us with three choices for complex wavefunctions: Either using Eq.~\eqref{eq:gradient_decompositon_highest_level} resulting in exact gradients for the exact fermionic generators with an associated cost factor of 4, or, using Eq.~\eqref{eq:gradient_decompositon_real_wfn} to approximate the gradient with an associated cost factor of 2, or, approximate the fermionic generators with either $G_+$ or $G_-$ where the exact gradient of the approximated generator can be obtained with Eq.~\eqref{eq:shift_rule} with an associated cost factor of 2.

\subsection{Operator Construction}
In the previous sections we showed how to obtain gradients of fermionic excitations with an overall cost factor of 4 (Eq.~\eqref{eq:gradient_decompositon_highest_level}) or a cost factor of 2 (Eq.~\eqref{eq:gradient_decompositon_real_wfn}) for real wavefunctions. The derivation holds in general for operators with 3 distinct eigenvalues $\left\{-r, 0, +r\right\}$. In Eq.~\eqref{eq:split_G_Gpm}, the original generators where formally split into two self-inverse parts $G_\pm$ which are later recombined to result in a fermionic shift gate~\eqref{eq:fermionic_shift_gate}, combining the shifted fermionic excitation gate and an additional unitary generated by the nullspace projector $P_0$ of the fermionic generator. In order to construct the fermionic shift gate in Eq.~\eqref{eq:fermionic_shift_gate}, one only needs the nullspace projector $P_0$ along with the original generator $G$, which for fermionic excitations is given in Eq~\eqref{eq:Gn}. Note that the explicit construction of the $G_\pm$ generators is never necessary, but they might be used in alternative implementations of the fermionic shift gate. Given the original generator $G$ and it's nullspace projector $P_0$, the $G_\pm$ as well as the original eigenspace projectors $P_\pm$ can be constructed as
\begin{align}
    &G_\pm = G \pm P_0 \\
    &P_\pm = \frac{1}{2}\left( G_\pm \pm 1 \right)\label{eq:Ppm_decomposition},
\end{align} 
where the completeness of the three eigenspace projectors was used.
In the following we will show how the corresponding nullspace projector can be constructed for fermionic excitation generators.
We will start with an intuitive illustration in the qubit representation and give the fermionic construction afterwards.

\subsubsection{Qubit Perspective}
In the Jordan-Wigner encoding $N$ spin-orbitals are directly mapped to $N$ qubits. The computational basis-states of the $N$ qubits correspond directly to the occupation number vectors in second-quantization and annihilation/creation operators are mapped to $\sigma_\pm = \frac{1}{2}\left(\sigma_x \pm i\sigma_y\right)$ qubit operators as well as $\sigma_Z$ operators on other qubits
\begin{align}
    a_k =& 1^{\otimes k-1} \sigma_k^+ \sigma_Z^{\otimes N-k}, \\
    a_k^\dagger =& 1^{\otimes k-1} \sigma_k^- \sigma_Z^{\otimes N-k}.\label{eq:JW}
\end{align}
The transformed generators~\eqref{eq:Gn} of fermionic excitations are acting with $\sigma_\pm$ operations onto qubits where electrons are excited from/to. 
Leaving potential phase changes introduced by $\sigma_z$ operators aside, they are acting as
\begin{align}
    \tilde{G}_{\mathbf{p}\mathbf{q}} = i\ket{o_\mathbf{p}^\mathbf{q}}\bra{o_\mathbf{q}^\mathbf{p}} - i\ket{o_\mathbf{q}^\mathbf{p}}\bra{o_\mathbf{p}^\mathbf{q}}\label{eq:G_qubit_approx}
\end{align}
where $o_\mathbf{p}^\mathbf{q}$ denotes all configurations where the $\mathbf{p}$ orbitals are occupied and the $\mathbf{q}$ orbitals are unoccupied with all combinations allowed for all other orbitals.
Note that the operators in Eq.~\eqref{eq:G_qubit_approx} neglect possible $\sigma_Z$ operations between the indexed qubits (that would, for example, be present in the single excitations of Eq.~\eqref{eq:G1}), those neglected operators will however not interfere with the structure of the nullspace and we will ignore them here for the sake of readability.
The nullspace of those generators is formed by the collection of configurations which are neither denoted by $o_\mathbf{p}^\mathbf{q}$ nor $o_\mathbf{q}^\mathbf{p}$. A convenient way to construct the nullspace is over it's complement, which is given by all configuration onto which the generator acts non-trivial. The projector $1-P_0$ onto the complement of the nullspace can then be constructed directly and rearranging leads to the expression for the nullspace projector as
\begin{align}
    \tilde{P}_{0;\mathbf{p}\mathbf{q}} = 1 - \ket{o_\mathbf{p}^\mathbf{q}}\bra{o_\mathbf{p}^\mathbf{q}} - \ket{o_\mathbf{q}^\mathbf{p}}\bra{o_\mathbf{q}^\mathbf{p}}.\label{eq:P0_qubit}
\end{align}
As an example, consider a two electron excitation generator that excites two electrons from the same spatial orbital to another spatial orbital, in a four spin-orbital representation this generator is $G=i\ket{0011}\bra{1100} - i\ket{1100}\bra{0011}$ with the nullspace projector $\tilde{P}_0= 1 - \ket{1100}\bra{1100} - \ket{0011}\bra{1100} = 1 - Q_-\otimes Q_- \otimes Q_+ \otimes Q_+ - Q_+\otimes Q_+ \otimes Q_- \otimes Q_-$, and $Q_\pm = \frac{1 \pm \sigma_z}{2}$.
Within a larger basis, there will be more qubits, but the generator will act trivial (or introducing a phase) on them, leading to all possible combinations in the nullspace projector $\sum_{xy\dots}\ket{1100xy\dots}\bra{1100xy\dots}$ that will sum up to unit operations as
$\sum_{xy\dots}\ket{xy\dots}\bra{xy\dots} = 1$ ending up in the same expression as before.
In general, the nullspace projector for an $n$-fold excitation can be constructed as
\begin{align}
    \tilde{P}_{0;\mathbf{p}\mathbf{q}} = 1 - \prod_{i=1}^n\left( Q_-(p_i)Q_+(q_i) + Q_+(p_i)Q_-(q_i) \right).\label{eq:P0_qubit_Qform}
\end{align}

\subsubsection{Fermionic Perspective}
Constructing all generators directly in their fermionic representation, will result in automatic gradient evaluation schemes independent of the underlying qubit encoding, allowing flexible adaption of new encodings~\cite{setia2018, chien2020custom} and improved compiling strategies into quantum gates~\cite{yordanov2020efficient}.
In analogy to the qubit construction in Eq.~\eqref{eq:P0_qubit} the nullspace projector of a general single, double or $n$-fold fermionic excitation~\eqref{eq:Gn} can be constructed as
\begin{align}
    P_{0;pq} =& 1 - N_p\tilde{N}_q - N_q\tilde{N}_p \\
    P_{0;pqrs} =& 1 - N_p\tilde{N}_qN_r\tilde{N}_s - N_s\tilde{N}_rN_q\tilde{N}_p \\
    P_{0;\mathbf{p}\mathbf{q}} =& 1 - N_{p_0}\tilde{N}_{q_0}\dots N_{p_n}\tilde{N}_{q_n} \nonumber\\ &- N_{q_0}\tilde{N}_{p_0}\dots N_{q_n}\tilde{N}_{p_n}.\label{eq:P0_fermionic}
\end{align}
using fermionic particle and hole number operators $N_{pq}=a^\dagger_pa_q$ and $\tilde{N}_{pq} = 1-N_{pq} = a_pa^\dagger_q$.
Note that using the Jordan-Wigner transformation~\eqref{eq:JW}, those operators get transformed into $Q_\pm$ (see also the next section) and Eq.~\eqref{eq:P0_fermionic} gets transformed into Eq.~\eqref{eq:P0_qubit_Qform}.

\subsection{Additional Cost in Quantum Gates}
Compared to the original circuit, an implementation of the fermionic shift gate~\eqref{eq:fermionic_shift_gate} will require additional gates, resulting from the $U_0^\alpha$ unitary that is generated by the nullspace projector of the fermionic generator. The individual cost of additional native quantum gates will depend on the qubit encoding of the fermionic algebra and individual properties of the underlying hardware, like their connectivity and native operations. See, for example, Ref.~\cite{lee2018generalized} for an estimate of the resources required for different variants of unitary coupled-cluster. It can however be expected, that those details will affect the $U_0^\alpha$ unitary similar ways as the other fermionic unitaries. We will do a first estimate by analysing the $P_0$ projector~\eqref{eq:P0_fermionic} in the Jordan-Wigner encoding~\eqref{eq:JW} given in Eq.~\eqref{eq:P0_qubit_Qform}.
and consider the $P_0$ projectors resulting from single and double excitations explicitly.
For the single excitation $P_0$ projector we get the encoding
\begin{align}
    \tilde{P}_{0;pq} &= 1 - Q_{-,p}Q_{+,q} - Q_{-,q}Q_{+,p}\nonumber\\ &= 1 - \frac{1}{2}\left(1-\sigma_z(p)\sigma_z(q)\right).
\end{align}
The $U_0^\alpha$ gate for single excitations can then be implemented as a two qubit gate generated by $\sigma_z(p)\sigma_z(q)$.
In the same way, the $P_0$ corresponding to a doubles excitation will result into 6 non trivial Pauli strings consisting of two $\sigma_z$ operations and one assembled from four $\sigma_z$ operations. Similar to the original fermionic generators the $P_0$ projector of an $n$-fold fermionic excitation will decompose into $\mathcal{O}\left(2^{2n-1}\right)$ individual Pauli strings which are in this case build up solely from $\sigma_z$ operations with the largest one having $\sigma_z$ on all $2n$ qubits. The gate cost for implementing the $U_0^\alpha$ unitary will therefore always be cheaper than for the associated fermionic excitation and we can upper bound it by the cost of implementing those.

\section{Applications and Examples}\label{sec:applications}
In the following we will illustrate potential initial applications for automatically differentiable unitary coupled-cluster. Note, that the techniques developed in this work allow convenient implementation of those techniques with computationally cheap gradients but can of course not guarantee that gradient based optimization schemes converge. The proposed applications are initial demonstrations of the automatic differentiation techniques developed within this work applied to specific examples and should not be viewed as benchmarks or fully defined methods. Our aim is rather to provide a generalized toolbox for automatic differentiable unitary coupled-cluster employable for the development of new methods. Our implementation is available within the free to use and open-source \textsc{tequila}~\cite{tequila} package. The improved gradient evaluation schemes are automatically applied to already implemented methods like UpCCGSD~\cite{lee2018generalized} that was employed in previously developed basis-set-free methods~\cite{kottmann2020reducing} and as an application for the meta-VQE~\cite{cerveralierta2020metavariational} approach. Apart from already existing implementations, the developed schemes can be employed for the development of new unitary coupled-cluster approaches in a blackboard style fashion. In the next sections we will illustrate this with explicit examples.

\subsection{Using the implementation within \textsc{tequila}}
The techniques developed within this work are implemented in the free and open-source package \textsc{tequila}~\cite{tequila} that operates on abstract expectation values of quantum circuits and operators, which can themselves be transformed and combined in an intuitive black-board style way. Our implementation is inspired by various open-source packages such as \textsc{madness}~\cite{harrison2016madness}, \textsc{pennylane}~\cite{bergholm2018pennylane}, and \textsc{diffiqult}~\cite{tamayo2018}, that are not focused on unitary coupled-cluster, but offer intuitive application programming interfaces, exposing, often highly specialized, numerical algorithms to a broad audience of interested scientists. \textsc{tequila} leverages state of the art high-performance simulators~\cite{qulacs, efthymiou2020qibo}, quantum chemistry software~\cite{openfermion, psi4, harrison2016madness} and \textsc{jax}~\cite{jax} for extended automatic differentiation techniques. Given that access rights are permitted, \textsc{tequila} can furthermore access state of the art quantum computers. Note however, that the quantum circuits of this work are still too deep in order to produce accurate results on current day hardware. Combined with the qubit-compressed, low-depth approaches of Ref.~\cite{kottmann2020reducing} successful demonstration on emerging quantum computers can however be anticipated. For the results of this work we interfaced \textsc{qulacs}~\cite{qulacs} as simulation backend, the BFGS implementation of \textsc{scipy}~\cite{scipy}, molecular integrals from \textsc{psi4}~\cite{psi42020} and qubit encodings from \textsc{openfermion}~\cite{openfermion}.\\

As an illustration of how our implementation can be employed we will give the explicit code that sequentially solves for ground and excited states of the Hydrogen molecule in a minimal representation where we have two electrons in four spin-orbitals - STO-3G(2,4). The results in the following sections are obtained in similar ways. The ground and excited states are solved using Eq.~\eqref{eq:excited_state_objective} and, in the explicit code example, we restrict the unitary to a single double excitation generated by $G_{(0,2),(1,3)}$ where two electrons are transferred between the two spin-up orbitals ($0$ and $2$) and spin-down orbitals ($1$ and $3$).

\begin{figure}[htbp]
    \centering
    \includegraphics[width=0.45\textwidth]{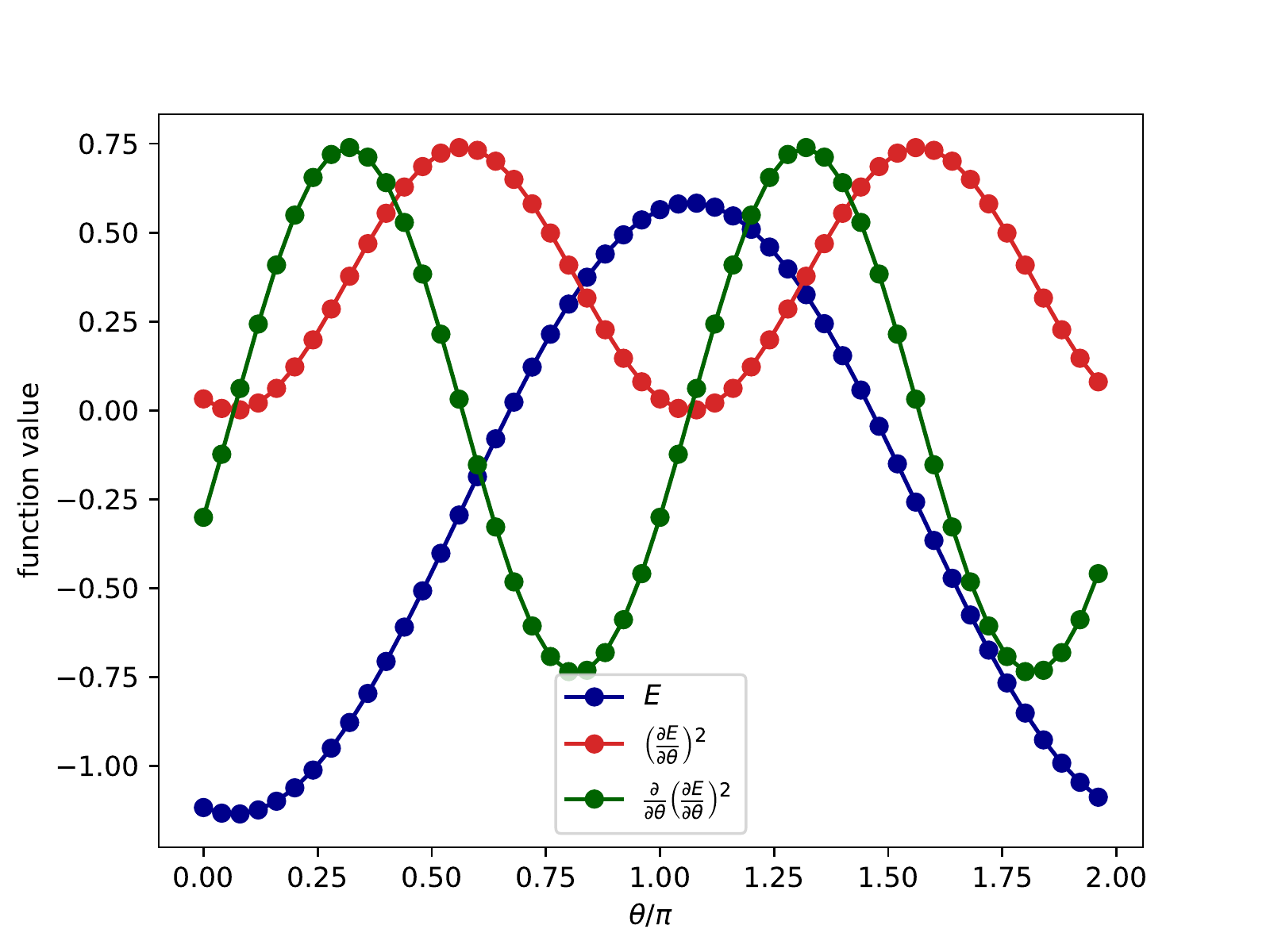}
    \caption{\textbf{Energy and gradients of a toy model:} Expectation value and gradients of the Hydrogen molecule in a minimal representation with a unitary restricted to a single double excitation. See the main text for explicit illustration of how to construct the abstract expectation values and objectives with \textsc{tequila}.}
    \label{fig:h2_example_pes}
\end{figure}

\begin{lstlisting}[language=Python]
import tequila as tq

# initialize molecular data
geom = "H 0.0 0.0 0.0\nH 0.0 0.0 0.7"
mol = tq.chemistry.Molecule(geometry=geom,
    basis_set="sto-3g",
    transformation="jordan_wigner")
    
H = mol.make_hamiltonian()

# index pairs for the double excitation
idx = [(0,2),(1,3)]

# construct ground-state objective
U0 = mol.prepare_reference()
U0 += mol.make_excitation_gate(idx, angle="a")
E0 = tq.ExpectationValue(H=H, U=U0)

# evaluate the objective at point 1.0
energy = tq.simulate(E0, {"a":1.0})
# compile and use as abstract function
energy_function = tq.compile(E0)
energy = energy_function({"a":1.0})

# optimize ground-state objective
result = tq.minimize("bfgs", E0)
variables=result.variables
energy0 = result.energy

# construct excited-state objective
U1 = mol.prepare_reference()
U1 += mol.make_excitation_gate(idx, angle="b")

Qp = tq.paulis.Qp(qubit=H.qubits)
E1 = tq.ExpectationValue(H=H, U=U1)
S2 = tq.ExpectationValue(H=Qp,U=U1+U0.dagger())

E = E1 - energy0*S2

# optimize excited-state objective
variables["b"] = 0.0
result = tq.minimize(method="bfgs",
    objective=E,
    variables=["b"],
    initial_values=variables)
energy1 = result.energy
\end{lstlisting}

since usually unitary coupled-cluster wavefunctions are real, the default gradient evaluation is done according to  Eq.~\eqref{eq:gradient_decompositon_real_wfn}. For complex wavefunctions the exact gradients can be demanded through further keywords in the \texttt{make\_excitation\_gate} function. As our explicit code example illustrates, the gradient evaluation procedures are handled automatically by \textsc{tequila}.
Both states could also be optimized by minimizing the square gradient of the energy expectation value $\min_{\theta} \left(\frac{\partial \expvals{H}{U(\theta)}}{\partial \theta}\right)^2$, a simplified strategy inspired by recent developments in classical mean-field theory~\cite{shea2020}. Using structures from the code block above, the \textsc{tequila} code block to directly deal with gradients of objectives looks like

\begin{lstlisting}[language=Python]
E = tq.ExpectationValue(U=U0, H=H)
dE = tq.grad(E, "a")
dE2 = dE**2
\end{lstlisting}

and the \texttt{dE2} objective can be used within the \texttt{minimize} function in the same way as illustrated above. The optimization will then converge to the ground or excited state depending on the optimization method and initial values. Note that gradient based optimization methods will then actually evaluate the gradient of $\left(\frac{\partial \expvals{H}{U(\theta)}}{\partial \theta}\right)^2$ without further specifications necessary. The abstract objective for this gradient could however be obtained in the same way as before using the \texttt{tq.grad} operation. In Fig.~\ref{fig:h2_example_pes} we plot the surfaces of those objectives explicitly. In the example code above we have illustrated how to evaluate the abstract objectives at specific angles. More details can be found in the online tutorials of \textsc{tequila}~\cite{tequila}. Note that the direct minimization of the square of the gradient will in general not result in accurate results for excited state calculations due to the presence of flat plateaus in the parameterized objective~\cite{mcclean2018barren}. More sophisticated optimization protocols are necessary to develop robust methods for this task providing good initial guesses and leveraging more generalized objective functions. It works for this basic illustration because of the simple shape of the one dimensional potential energy surface (see Fig.~\ref{fig:h2_example_pes}) and we are using it as a first example where the gradient directly enters the objective function in the hope that in can be the first step for future method development.

\subsection{An Illustrative Example}
\begin{figure*}[htbp]
    \centering
    \includegraphics[width=0.4\textwidth]{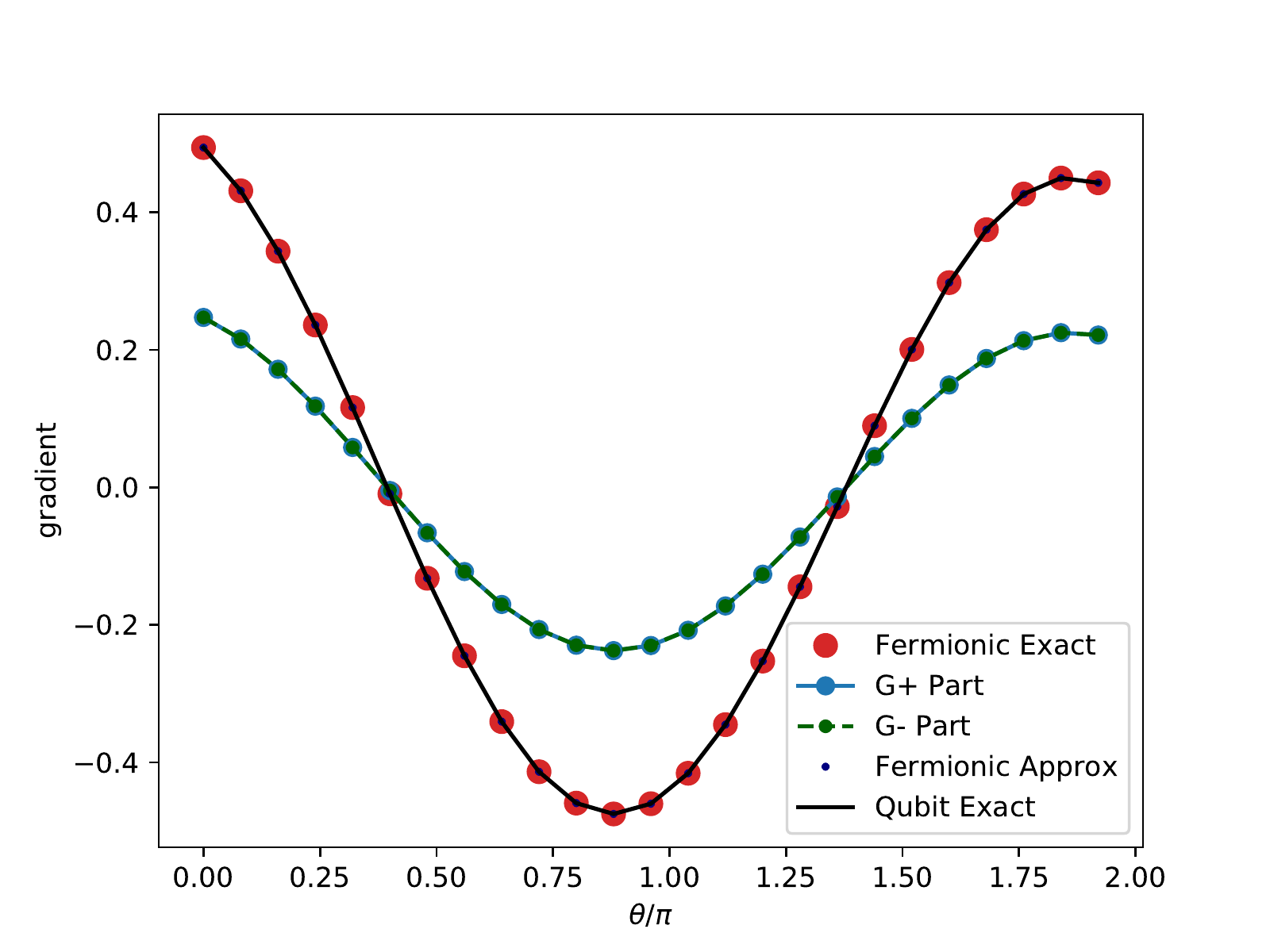}
    \includegraphics[width=0.4\textwidth]{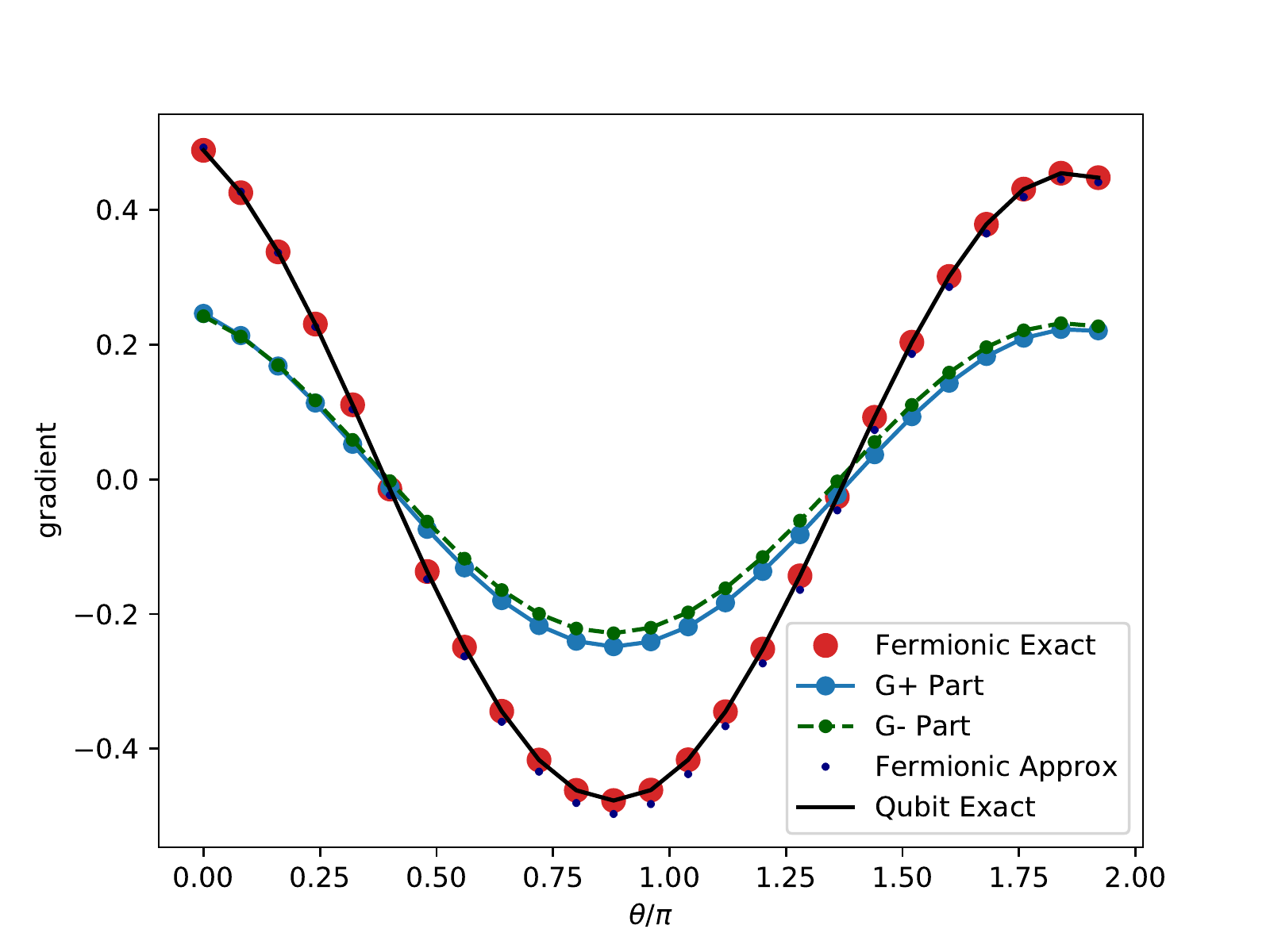}
    \caption{\textbf{An Illustrative Example:} We show the gradients of the energy expectation value for the Hydrogen molecule in 4 spatial orbitals (6-31G) with respect to a specific generator where the wavefunctions are prepared in a way that ensures overlap with the nullspace of the generator (see main text). The underlying wavefunctions are real (left) and complex (right) and the gradients are computed according to Eq.~\eqref{eq:gradient_decompositon_highest_level} (fermionic exact, Fig.~\ref{fig:algorithm_overview} middle) where the two parts of the sum ($G_+$ and $G_-$ corresponding to $\alpha \in \left\{ +1, -1 \right\}$) are also shown in the plots. The approximated fermionic gradient (Fig.~\ref{fig:algorithm_overview} bottom) is according to Eq.~\eqref{eq:gradient_decompositon_real_wfn}. The exact gradient in the qubit representation is computed by automatic differentiation of the individual rotational gates in the compiled circuit (Fig.~\ref{fig:algorithm_overview} top).}
    \label{fig:simple_illustration}
\end{figure*}

We will use a simple example to further illustrate some the concepts of the last section. In particular we will demonstrate that the simplified gradient evaluation scheme of Eq.~\eqref{eq:gradient_decompositon_real_wfn} is not exact for complex wavefunctions.
As a toy model we will use a specific circuit for a hydrogen molecule in 4 spatial orbitals (6-31G basis) - similar to the last section, just with more orbitals in the model.
Consider the fermionic excitation generator $G_{0213}$ that will excite electrons from configurations $\ket{0011\cdot\cdot\cdot\cdot}$ to $\ket{1100\cdot\cdot\cdot\cdot}$, where the $\cdot$ can be arbitrary combinations of $0$ and $1$. The nullspace of this generator is spanned by all configurations whose bitstring representations do not start with $0011$ or $1100$.
Consider the quantum circuit
\begin{align}
    &U = \sigma_x(0)\sigma_x(1)e^{-i\frac{1}{2}G_{0213}}e^{-i\frac{1}{2}G_{0411}}e^{-i\frac{1}{2}A_{0213}},
\end{align}
with $A \in \left\{ G, G_+ \right\}$.
This circuit will prepare a superposition of the configurations $\ket{11000000},\ket{00110000}$ and $\ket{00001100}$ and depending on the choice of the generator A the wavefunction will be real ($A=G$) or complex ($A=G_+$) since the unitary generated from $A=G_+$ will introduce a complex phase (see Eq.~\eqref{eq:Gpm_decomposition}) to all configurations that are part of its nullspace (here $\ket{00001100}$).
Now we will add a parameterized unitary generated by $G_{0213}$ and take the expectation value with respect to the electronic Hamiltonian
\begin{align}
    \expvals{H}{U(\theta)},\quad U(\theta) = U e^{-i\frac{\theta}{2}G_{0213}}.
\end{align}
The gradients of this expectation value computed with different methods are shown in Fig.~\ref{fig:simple_illustration}. For the complex wavefunction the approximate scheme where the two parts of the sum in Eq.~\eqref{eq:gradient_decompositon_highest_level} are considered to be equivalent leads to slight deviations. This individual contributions to the exact gradient are also shown in Fig.~\ref{fig:simple_illustration}. For the real wavefunction the individual contributions of $G_\pm$ are identical and the approximation becomes exact.

\begin{figure*}[htbp]
    \centering
    {\includegraphics[width=0.33\textwidth]{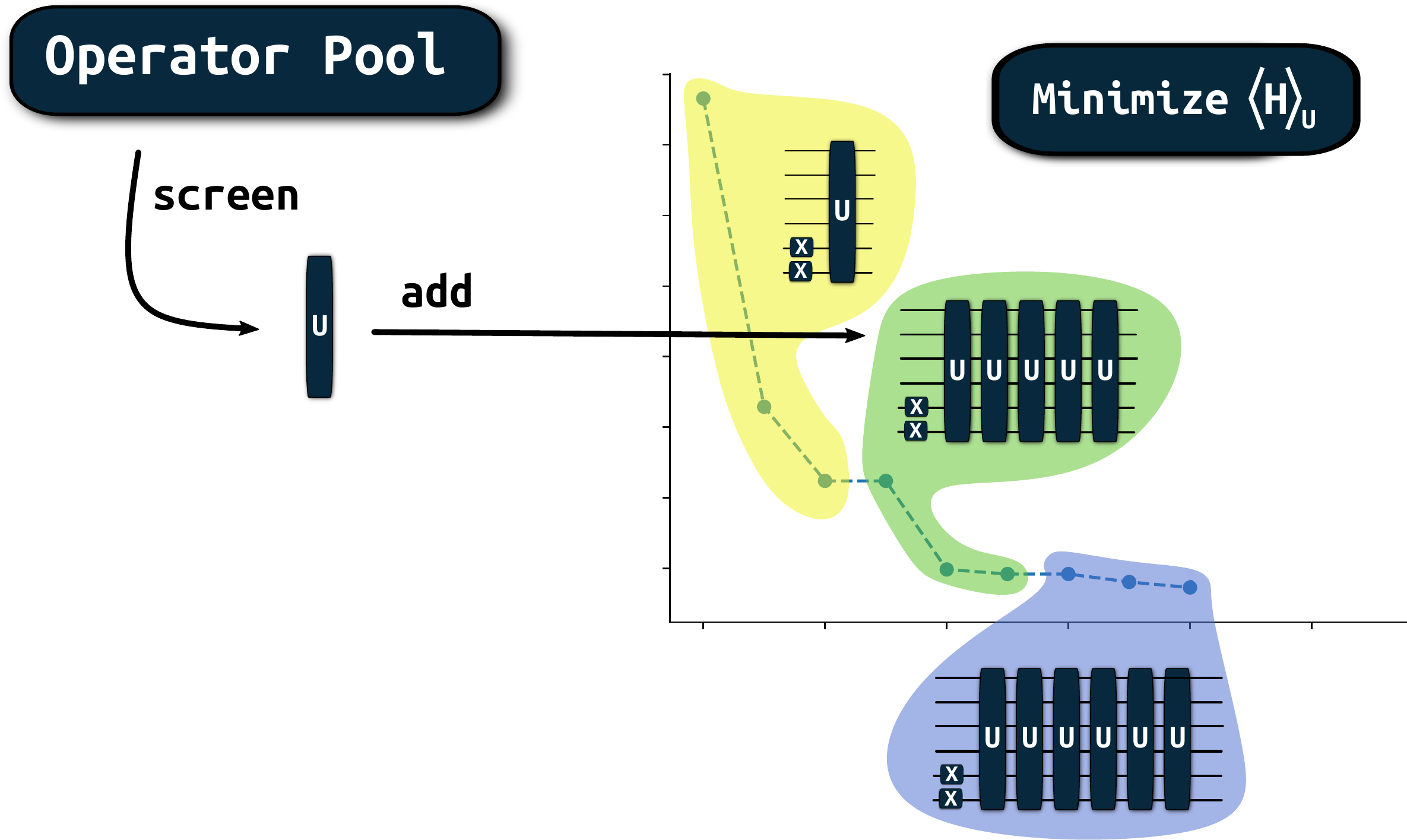}}
    {\includegraphics[width=0.3\textwidth]{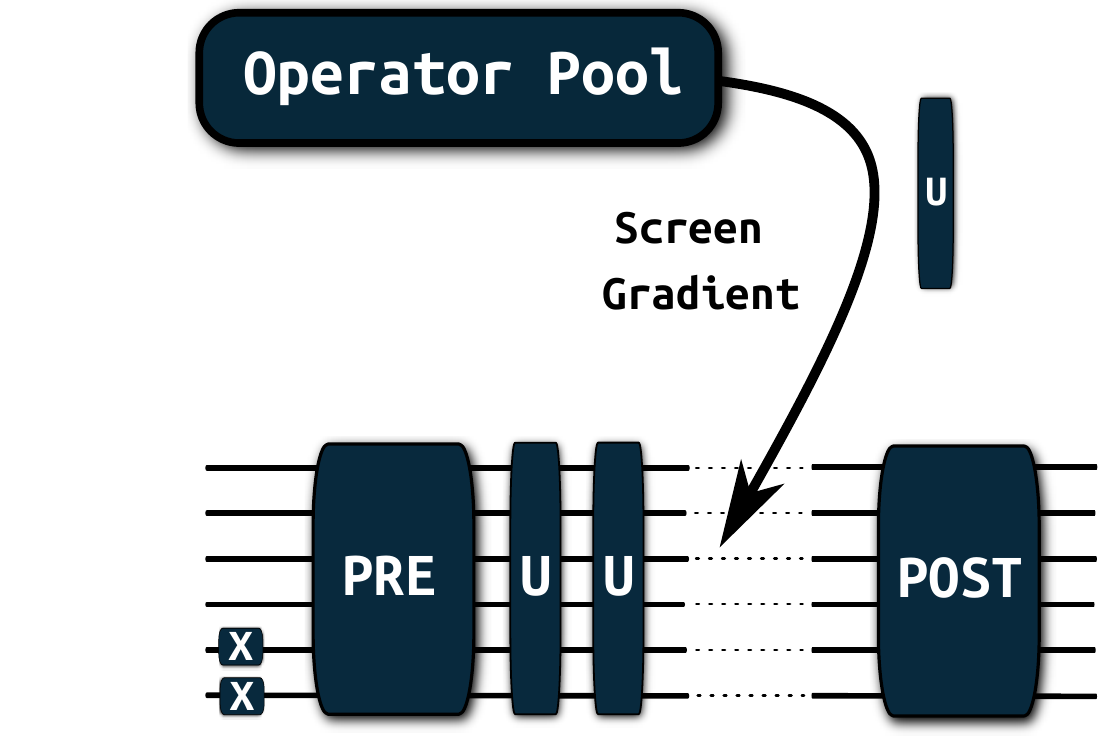}}\hphantom{aaaaa}
    {\includegraphics[width=0.18\textwidth]{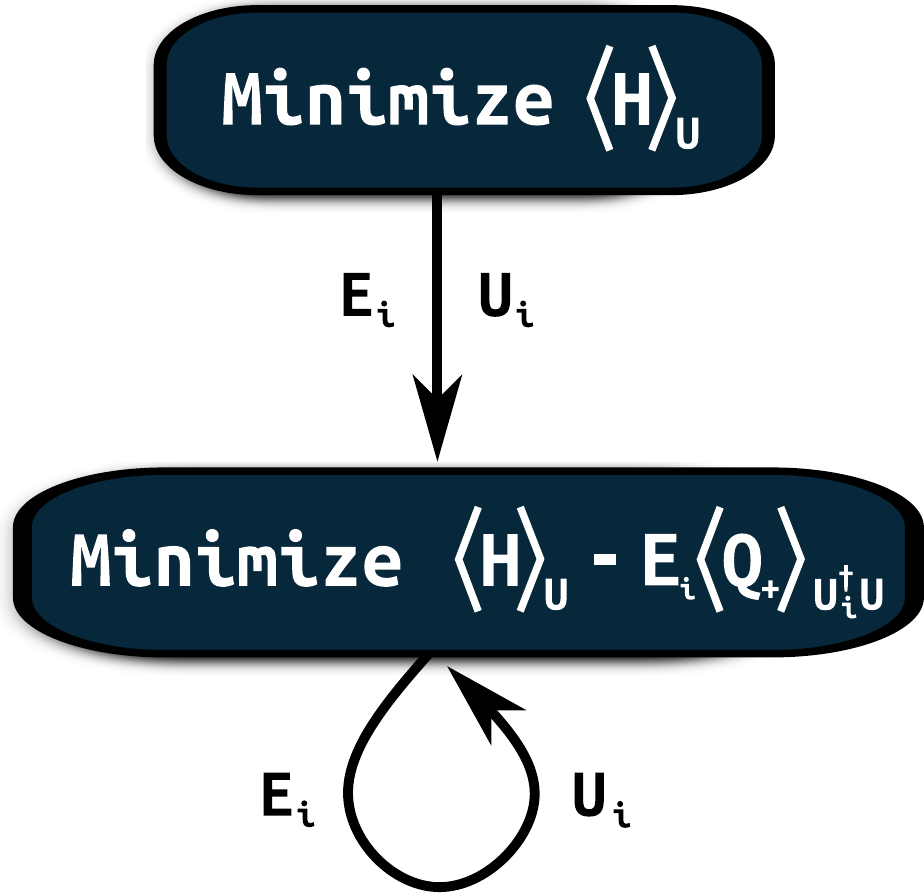}}
    \caption{\textbf{Initial applications for automatically differentiable unitary coupled-cluster:} Basic procedure for adaptive circuit construction (left). Generalized framework combining adaptive and static blocks (middle). Sequential solver for excited states (right).}
    \label{fig:applications_overview}
\end{figure*}

\begin{figure*}
    \centering
    \includegraphics[width=0.3\textwidth]{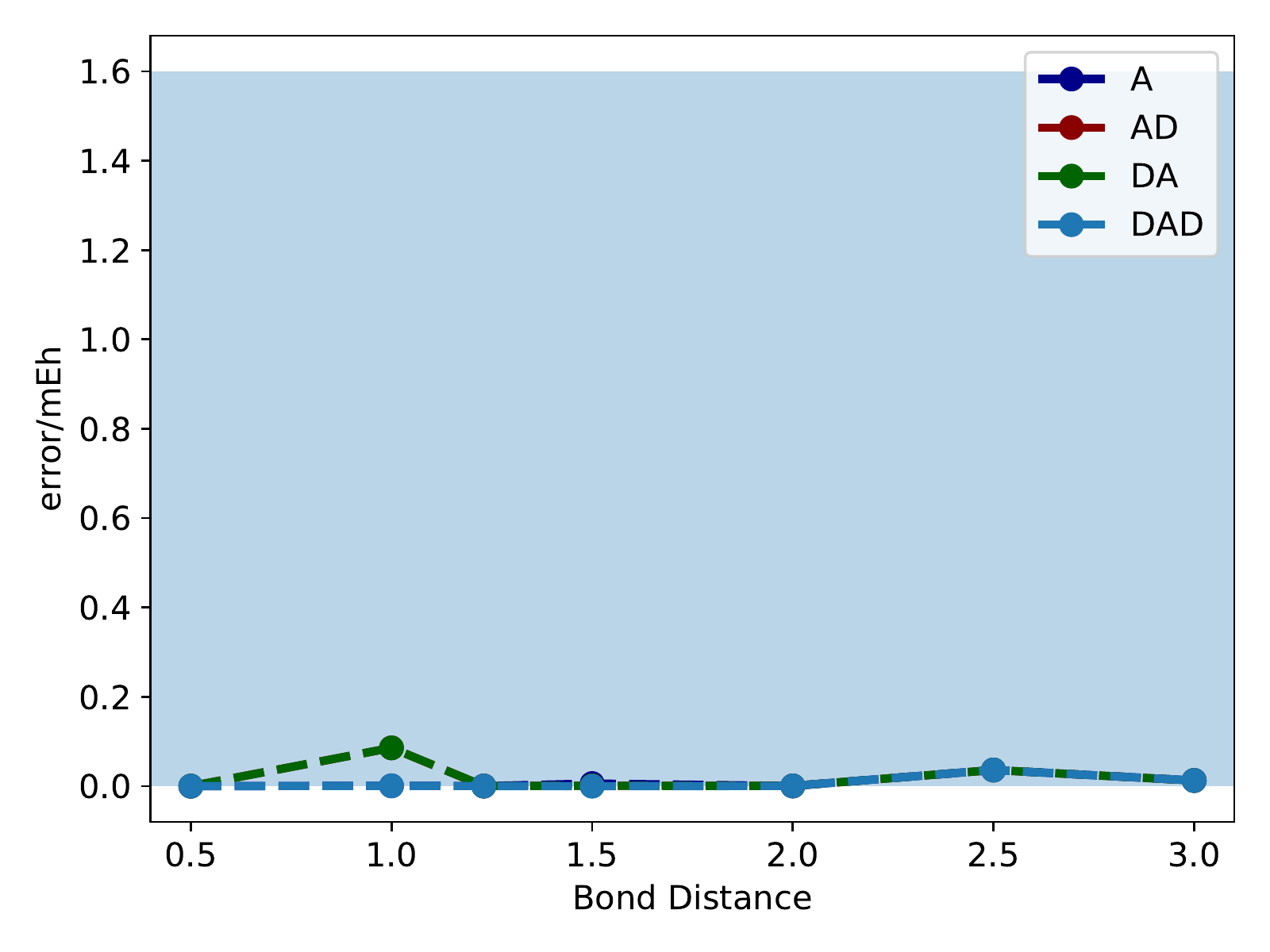}
    \includegraphics[width=0.3\textwidth]{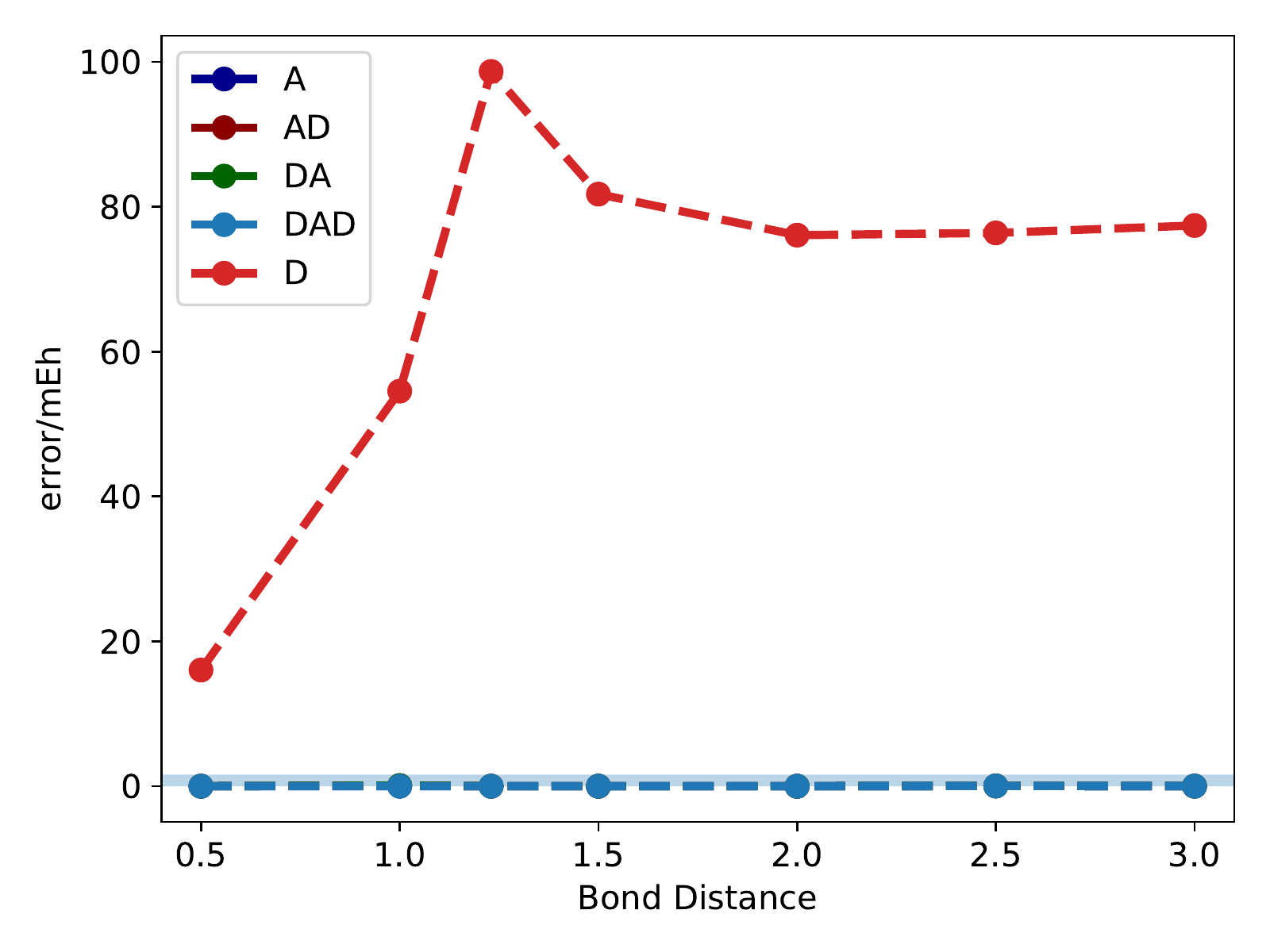}
    \includegraphics[width=0.3\textwidth]{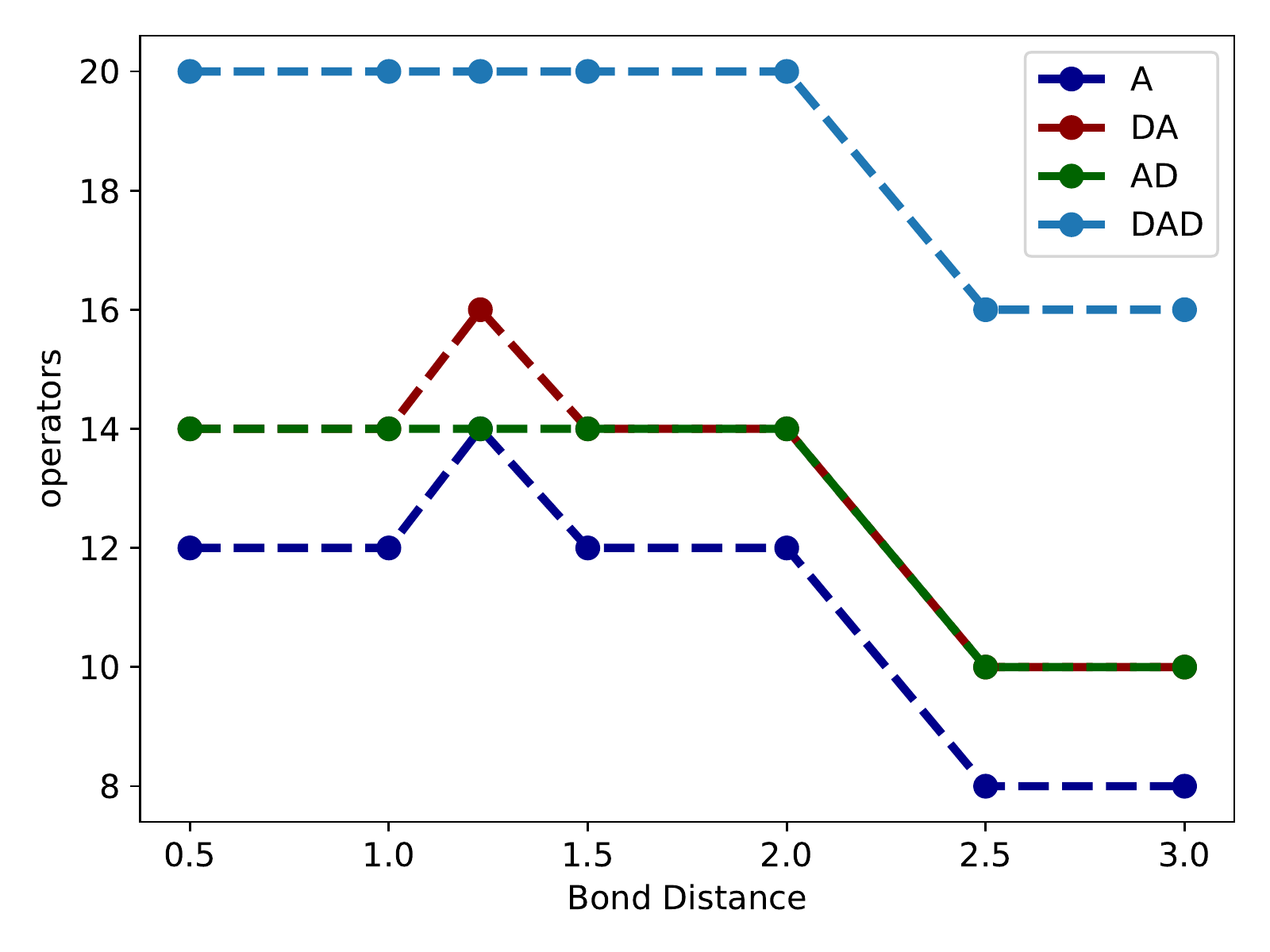}

    \includegraphics[width=0.3\textwidth]{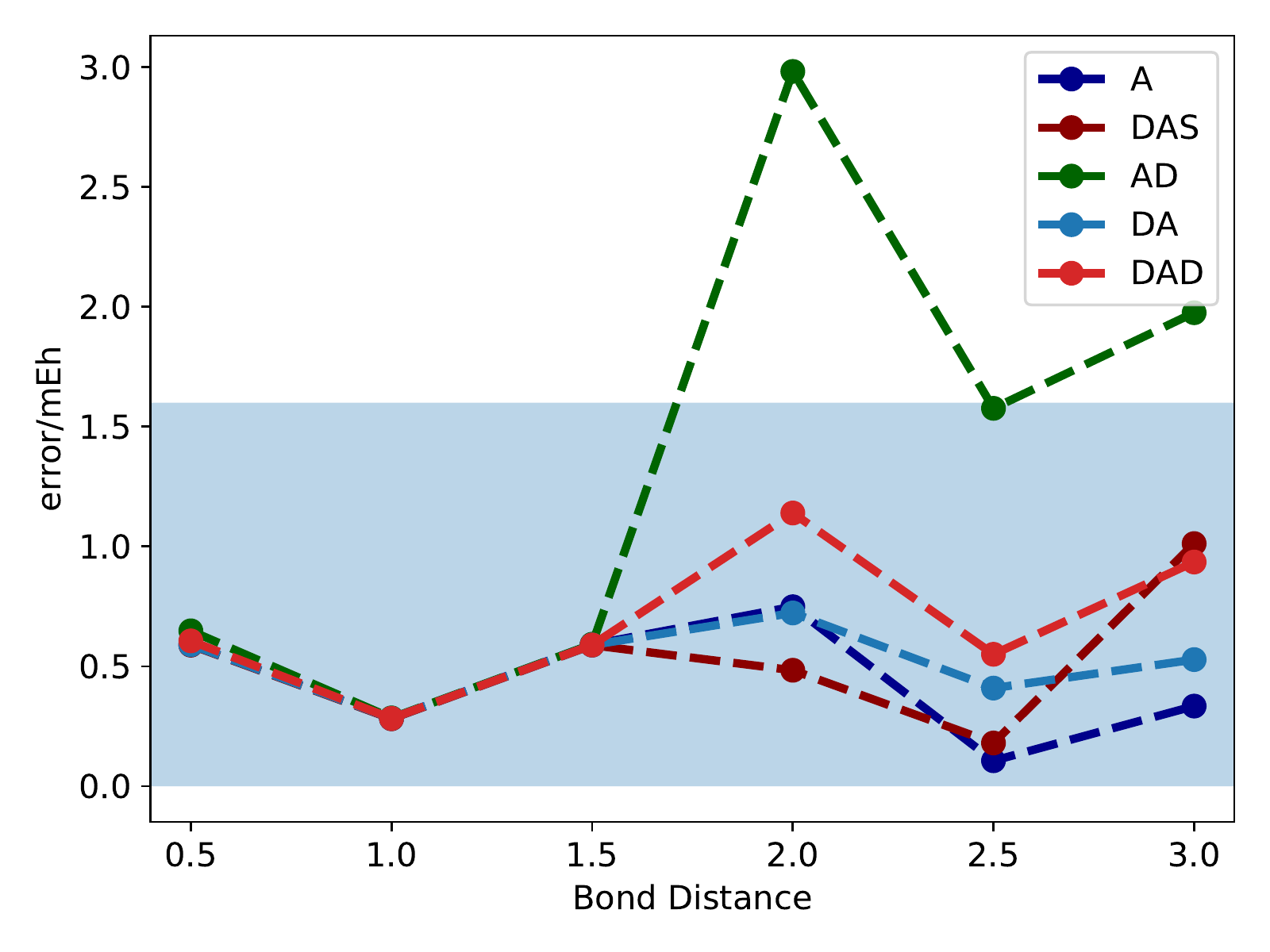}
    \includegraphics[width=0.3\textwidth]{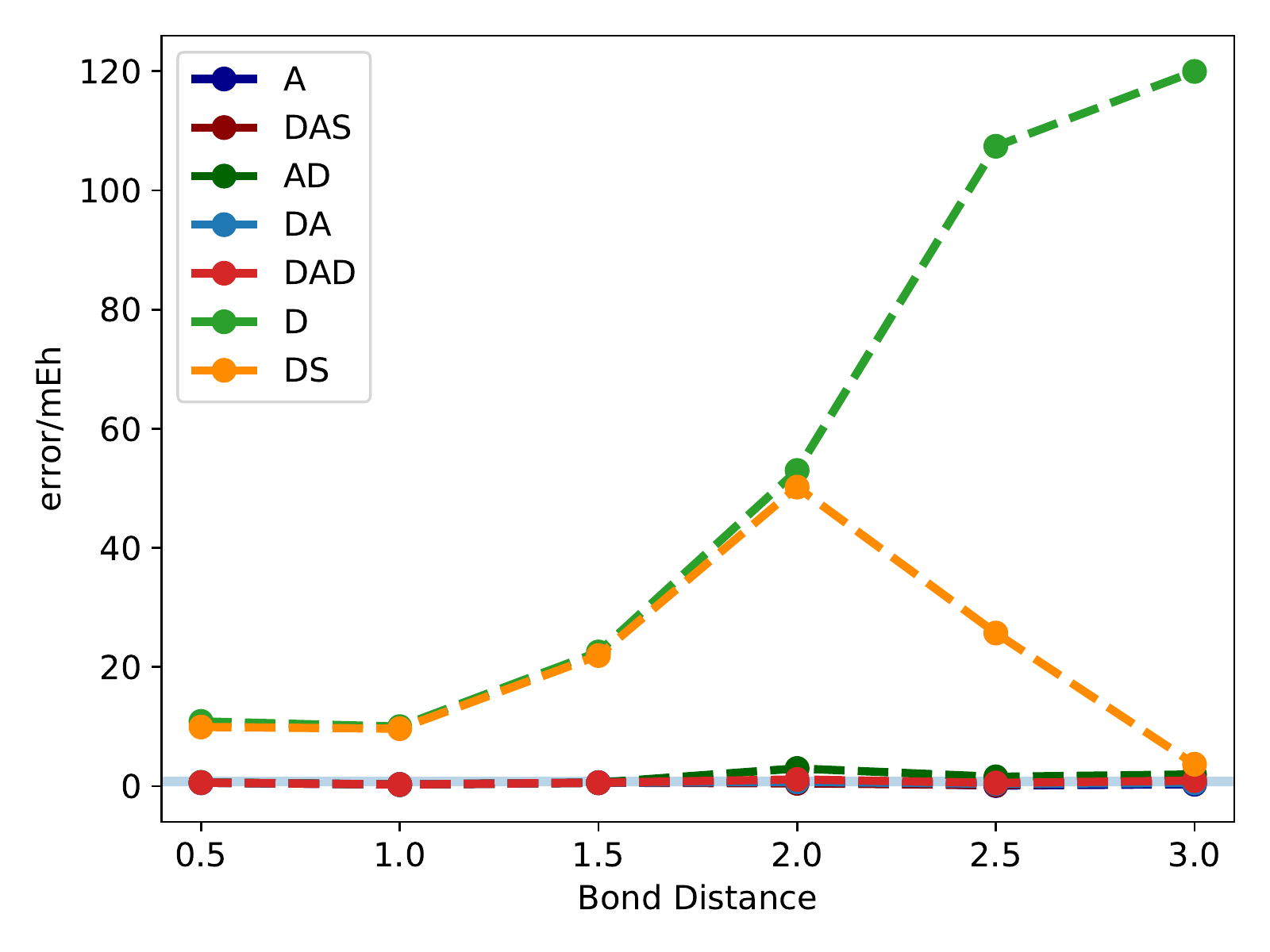}
    \includegraphics[width=0.3\textwidth]{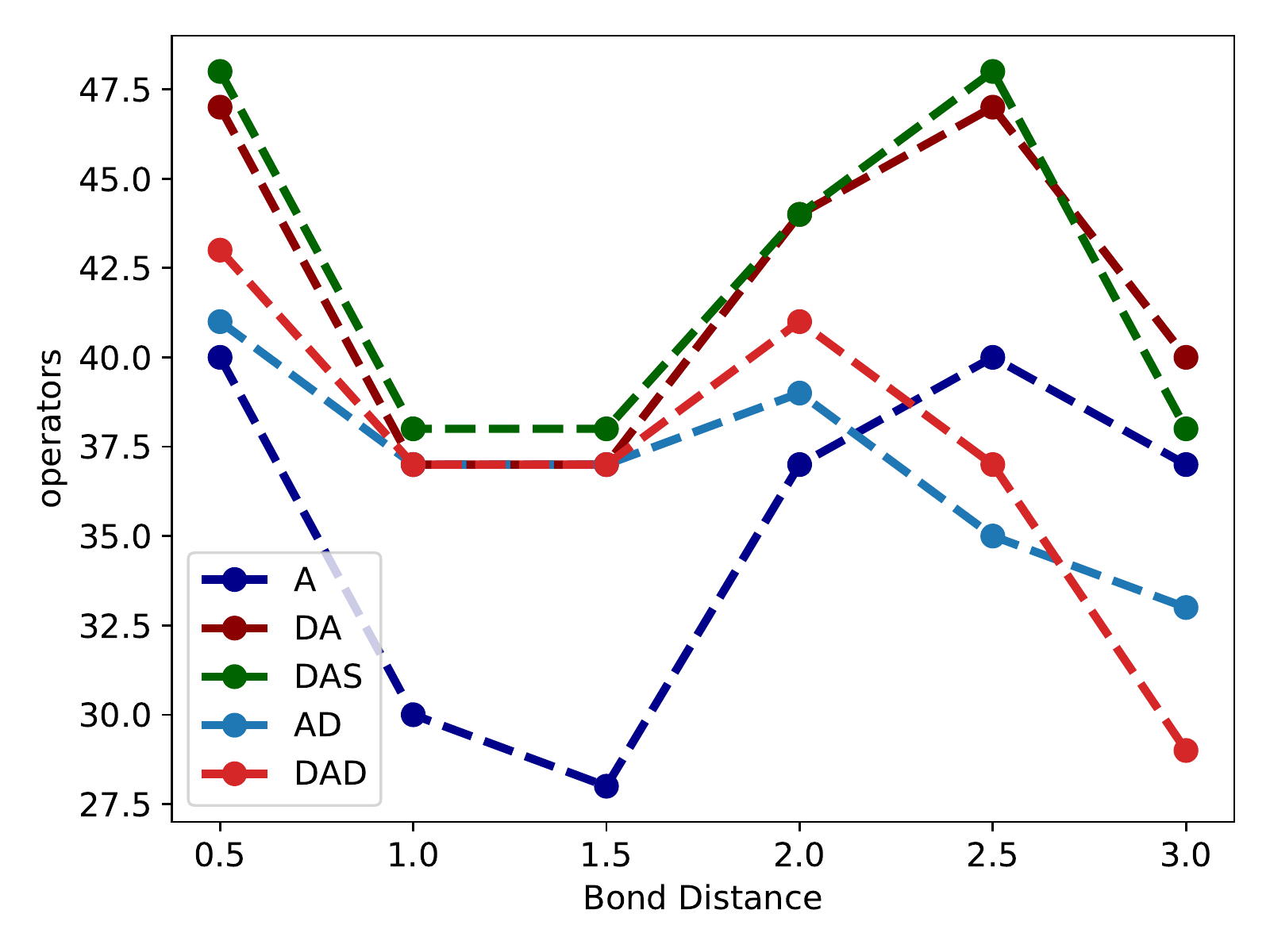}
    \caption{\textbf{Combining adaptive and static methods:} Combination of adaptively growing unitaries (A) with static blocks of generalized single (S) and double (D) excitations. Results are shown for H$_4$/STO-3G(4,8) (top) and BeH$_2$/STO-3G(6,14) (bottom).  We used generalized doubles restricted to pair excitations for the D blocks (6 operators for H$_4$ and 15 for BeH$_2$). For the DAD results the generalized doubles block was split up into standard UpCCD doubles and the residual set of operators. In the DAS approach the trailing doubles are replaced with generalized single excitations. Excitations involving the 1s orbital of Be were not included into the static blocks D and S. The plots in the center show in addition the results without the adaptive blocks.}
    \label{fig:adapt_vqe_results}
\end{figure*}

\subsection{Adapt VQE}
Adaptive approaches where the chain of excitations in the unitary is iteratively increased by adding operators from a operator pool based on gradient based screening processes (see Fig.~\ref{fig:applications_overview} for a high level overview).
These type of algorithms have been successfully applied in different flavors like qubit-coupled-cluster~\cite{ryabinkin2018qubit, ryabinkin2020iterative, lang2020iterative} or adapt-vqe~\cite{grimsley2019adaptive, tang2019qubit} which mostly differ in the way they screen and construct operators.
In these approaches the commutator between the Hamiltonian and the generator of potential excitations is used in the screening process to compute the gradient.
In order to perform the actual optimization with gradient based methods on a quantum computer the commutator approach would only work for the gradient of the operator added last to the unitary circuit.
With an automatically differentiable framework screening as well as optimization can be treated in the same way. This allows for more generalized adaptive growth procedures where the adaptive part is not restricted to be the trailing part of the quantum circuit. In Fig.~\ref{fig:adapt_vqe_results} we show some initial demonstrations combining static and adaptive blocks (see Fig.~\ref{fig:applications_overview} for an illustration) where we used a restricted set of fermionic single and double excitations for the static blocks where the double excitations are restricted to pair excitations in the same way as in the UpCCGSD approach~\cite{lee2018generalized}. As model systems we chose the H$_4$/STO-3G system as in Ref.~\cite{lee2018generalized} where we vary the distance between the individual H$_2$ molecules, and the BeH$_2$/STO-3G molecule, where we vary the two Be-H distances at the same time. By varying the corresponding distances we generate different problem instances that are representative of electronic structure problems without being too specialized.

\subsection{Excited Adapt VQE}
As another application we use a modified version of the adaptive ground-state algorithm in order to optimize excited states where we follow the strategy  applied in combination with the \textit{k-UpCCGSD}\cite{lee2018generalized} model of unitary coupled-cluster.
A variational quantum algorithm for bound excited states (states with negative energies) can be achieved sequentially by projecting out previously solved solutions (see Fig.~\ref{fig:applications_overview} for an illustration).
Since it is known how to prepare previously found solutions with the unitaries $U_i$ the variational preparation of the target excited state becomes equal to the minimization of
\begin{align}
    E = \langle H \rangle_{U\left(\theta\right)} - \sum_i E_i \langle \mathbf{Q_+} \rangle_{U_i^\dagger U\left(\theta\right)},\label{eq:excited_state_objective}
\end{align}
where $\langle O \rangle_{U}$ denotes the expectation value of the operator $O$ with respect to the wave function prepared by the unitary $U$ and the operator $\mathbf{Q_+}$ denotes the projector on the \textit{all-zero} state
\begin{align}
    \mathbf{Q_+} = \ket{0\dots0}\bra{0\dots0} = \bigotimes^N Q_+.
\end{align}
The second term of Eq.~\eqref{eq:excited_state_objective} computes the square of the overlap between the two wave functions, scaled by the energy of the previous state, and ensures orthogonality to all previously found solutions (see appendix). 
Note that unbound states can in principle also be found with this approach by replacing the energies with large positive factors.
Estimating overlaps like this is an alternative approach to the computationally costly SWAP test based strategies originally proposed for excited state solvers in Ref.~\cite{higgott2019variational} and successfully applied for similar systems as in this work by using imaginary time evolution in Ref.~\cite{jones2019}. 
The same strategy for overlap estimation could be applied successfully in the optimization of quantum optical setups~\cite{kottmann2020quantum}.
One important property of the $\mathbf{Q_+}$ operator is that, other than for the Hamiltonian $H$, all of its components naturally commute which allows sampling of all terms within a single run.
Compared to the Hamiltonian $H$ the additional measurements coming from $\mathbf{Q_+}$ are negligible.
\begin{figure*}
    \centering
    \includegraphics[width=0.4\textwidth]{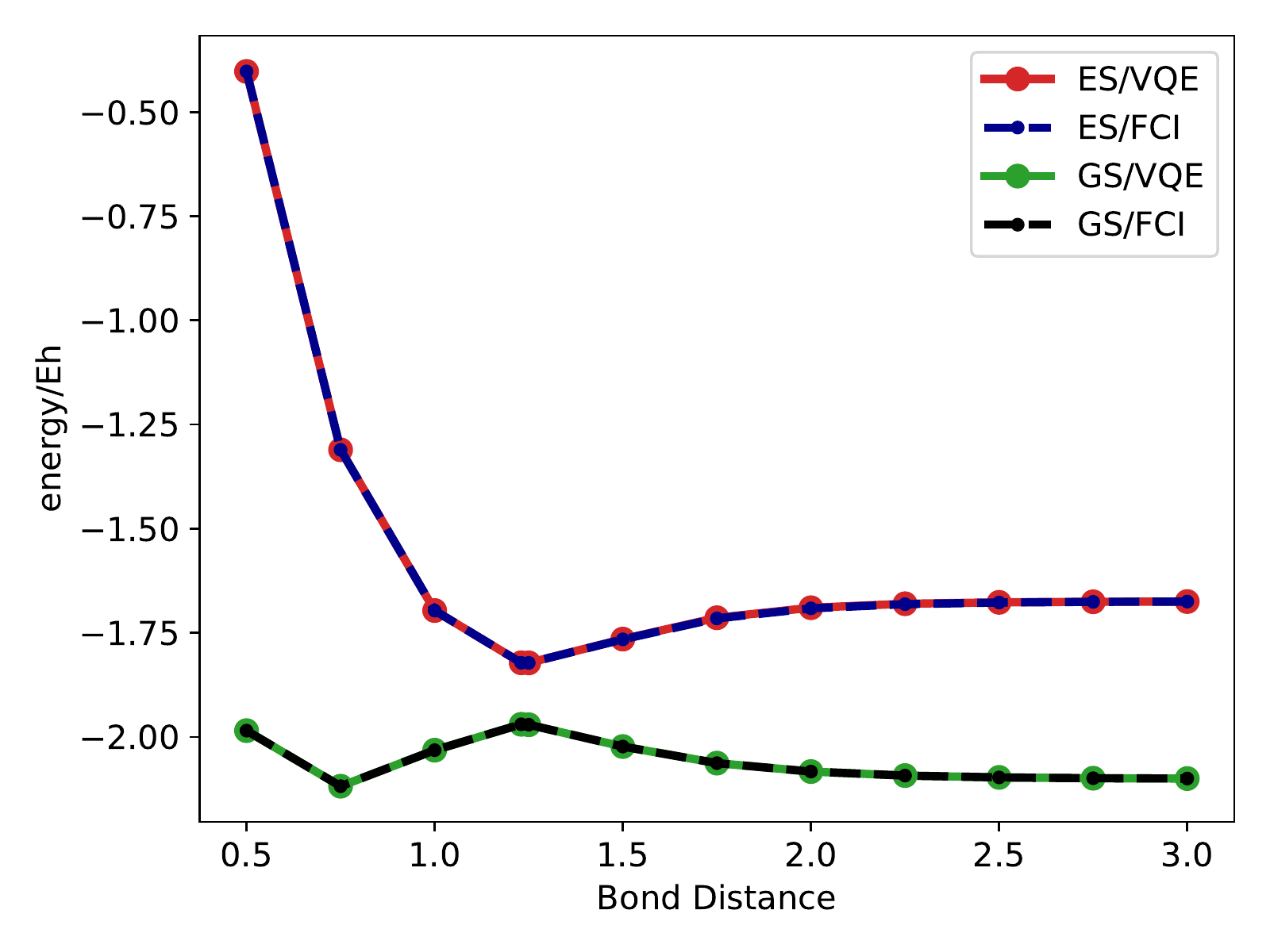}
    \includegraphics[width=0.4\textwidth]{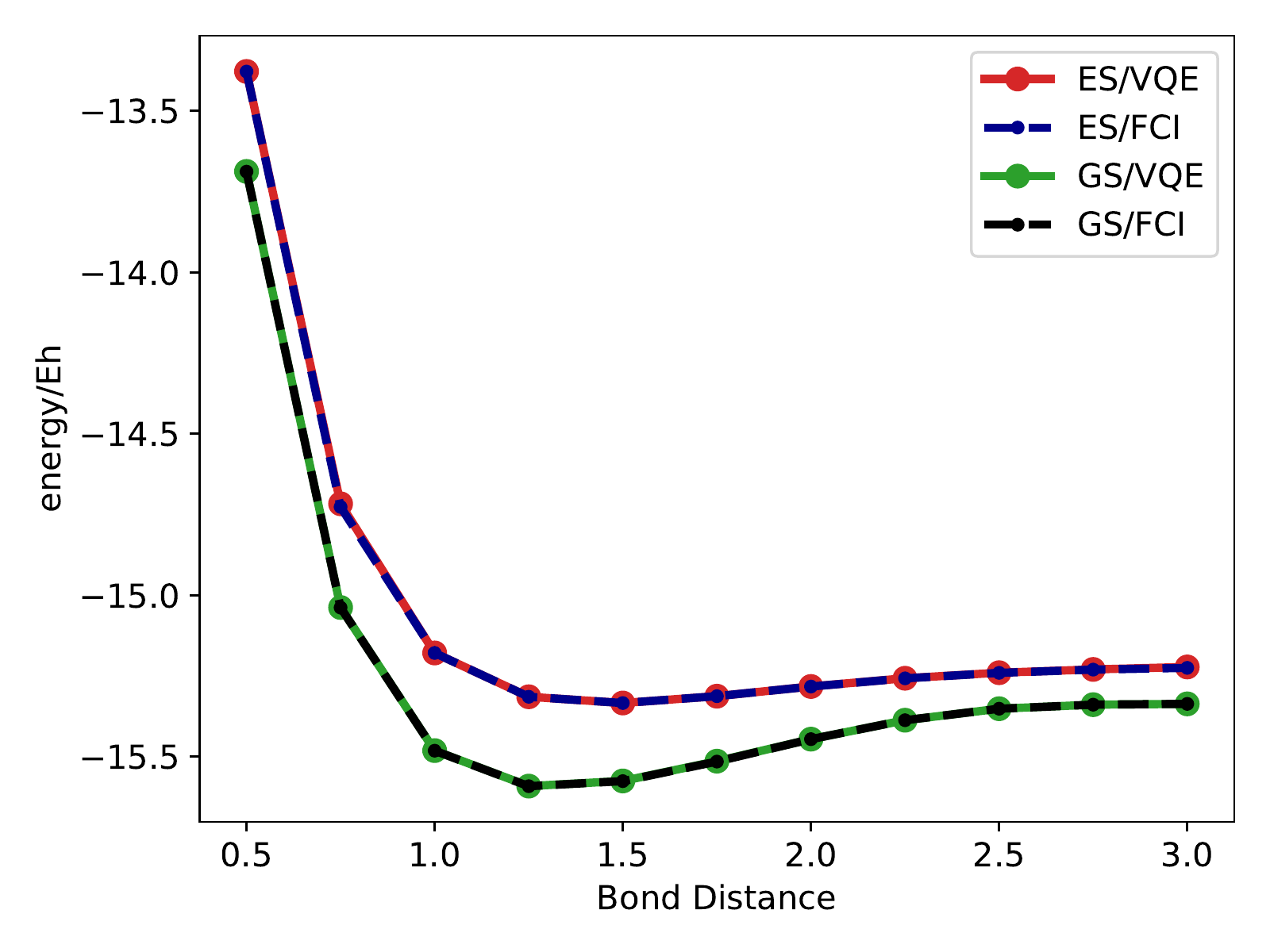}
    \caption{\textbf{Adapt-VQE for ground and excited states}: Adapt-VQE results using automatic differentiation for the screening and the optimization for ground and first excited state energies of H$_4$/STO-3G(4,8) (left) and BeH$_2$/STO-3G(6,14) (right). Except for the last point of the BeH$_2$ excited state, all points agree to millihartree accuracy with the corresponding exact solution (FCI) in the given basis set. See also Fig.~\ref{fig:applications_overview}. We included the special point at distance $1.23${\AA} with a square configuration.  }
    \label{fig:adapt_vqe_results_ex}
\end{figure*}
We combine the sequential strategy with adaptive solvers by simply replacing the original objective function, which was just the expectation value of the Hamiltonian, with Eq.~\eqref{eq:excited_state_objective} and solve sequentially for low lying excited states.
In Fig.~\ref{fig:adapt_vqe_results_ex} we show some results for H$_4$ and BeH$_2$. In the case of BeH$_2$ we used the dominant contribution of the lowest configuration interaction singles solution as reference for the excited state calculation (see the appendix for explicit circuits) and for H$_4$ we used the Hartree-Fock reference for both calculations.
The reason for this being that the lowest lying electronic singlet state of H$_4$ has the same irreducible representation as the ground-state which is not the case for BeH$_4$. So this particular excited state of BeH$_4$ could have also been found with a ground-state algorithm with a symmetry restricted pool of excitations.
In both cases, we restricted the adaptive optimization to stay within the singlet manifold, in order to avoid collapsing to the lower lying triplet states.
Note, that commutator based screening techniques are not possible in this case since the adaptive circuit is not the trailing part of the overlap expectation value.
In order to make them work, the $U_i^\dagger$ unitaries would have to be folded into the $P_0$ operator, increasing the number of measurements significantly (see \cite{kottmann2020quantum} for a similar argument).
Such folding techniques are achieved in an iterative version of qubit-coupled-cluster~\cite{ryabinkin2020iterative}, using the properties of self-inverse qubit generators. Equation~\eqref{eq:G_decomposition} could be employed to develop similar strategies in the fermionic representation.

\section{Conclusion and Outlook}\label{sec:conclude}
Variational algorithms are currently one of the most promising applications on current and future quantum computers.
Quantum chemistry is one of the target fields of those algorithms and expectations are high for new types of methods developed within that framework.
Analytical gradients for unitary coupled-cluster type approaches can in principle be computed on quantum computers, they come however with high computational costs using standard techniques on the qubit level.
We developed the necessary techniques in order to evaluate analytical gradients of general $n$-fold fermionic excitation operators with a cost factor of $4$ in general and factor of $2$ for real wavefunctions.
Our strategies to compile gradients can be done entirely in the fermionic representation making it independent of the used qubit mapping.
The developed techniques combined with \textsc{tequila}s automatic differentiation framework provide a testbed for quantum chemistry on quantum computers where new ideas, like low-depth approaches based on pair-natural orbitals~\cite{kottmann2020reducing} or Krylov subspaces~\cite{huggins2019nonorthogonal,stair2019multireference}, can be prototyped and demonstrated in a blackboard fashion. Our implementation provides an easy to use, automatically differentiable framework for unitary-coupled cluster, that leverages state of the art high performance simulators~\cite{efthymiou2020qibo, qulacs} and is ready for emerging quantum computers. 
We demonstrated initial applications for ground and excited state calculations for small model systems where we extended adaptive circuit construction schemes and, for the first time, applied them to excited state optimization.

\section{Acknowledgement}
We would like to thank Philipp Schleich and Sumner Alperin-Lea for providing valuable suggestions and comments on the manuscript.
A.A.-G. acknowledges the generous support from Google, Inc.  in the form of a Google Focused Award.
This work was supported by the U.S. Department of Energy under Award No. DE-SC0019374 and the U.S. Office of Naval Research (ONS-506661).
A.A.-G. also acknowledges support from the Canada Industrial Research Chairs  Program and the Canada 150 Research Chairs Program. Computations were performed on the niagara supercomputer at the SciNet HPC Consortium.~\cite{niagara1, niagara2} SciNet is funded by: the Canada Foundation for Innovation; the Government of Ontario; Ontario Research Fund - Research Excellence; and the University of Toronto.
We thank the generous support of Anders G. Fr\o{}seth.

\bibliography{main.bib}

\newpage
\onecolumngrid
\section*{Appendix}
\subsection*{Eigenstates of fermionic generators}
Take, without loss of generality, Eq.~\eqref{eq:G_qubit_approx} of the main text.
Following the definitions of the main text, the eigenstates of this operator can directly be constructed to be
\begin{align}
    \ket{\pm} = \frac{1}{\sqrt{2}}\left( \ket{o^\mathbf{p}_\mathbf{q}} \pm \ket{o^\mathbf{q}_\mathbf{p}}\right)
\end{align}
with corresponding eigenvalues $\pm 1$. On all configuration outside that manifold the generator will act as zero making them also eigenstates with eigenvalue $0$.

\subsection*{Derivation of Eq.~\eqref{eq:G_decomposition}}
\begin{align}
U\left(\theta\right) &= e^{-i\frac{\theta}{2}G} = e^{-i\frac{\theta}{4}\left(G_+ + G_-\right)} \\
&= e^{-i\frac{\theta}{4}G_+}e^{-i\frac{\theta}{4}G_-} \\
&= \left(\cos\left(\frac{\theta}{4}\right) - i \sin\left(\frac{\theta}{4}\right)G+\right) \left(\cos\left(\frac{\theta}{4}\right) - i \sin\left(\frac{\theta}{4}\right)G-\right)\\
&= \cos^2\left(\frac{\theta}{4}\right) - i \cos\left(\frac{\theta}{4}\right)\sin\left(\frac{\theta}{4}\right)\left(G_+ + G_-\right) - \sin^2\left(\frac{\theta}{4}\right)G_+G_-\\
&= \left(\cos^2\left(\frac{\theta}{4}\right)-\sin^2\left(\frac{\theta}{4}\right)\right) - 2i\cos\left(\frac{\theta}{4}\right)\sin\left(\frac{\theta}{4}\right)G + 2\sin^2\left(\frac{\theta}{4}\right)P_0\\
&=\cos\left(\frac{\theta}{2}\right) -i\sin\left(\frac{\theta}{2}\right)G + \left(1-\cos\left(\frac{\theta}{2}\right)\right)P_0
\end{align}
For the derivation we used the following properties and identities
\begin{align}
    &G = P_1 - P_{-1}\\
    &P_1 + P_{-1} + P_0 = 1\\
    &G^2 = P_1 + P_{-1} \\
    &P_iP_j = \delta_{ij}, \quad i,j\in\left\{-1,1,0\right\}\\
    &GP_0 = P_0G = 0\\
    &G_+G_- = G_-G_+ = G^2 - P_0 = 1-2P_0\\
    &\left[G_+, G_-\right] = 0\\
    &G_+^2 = G_-^2 = 1\\
    &\left(\cos^2\left(\frac{\theta}{4}\right)-\sin^2\left(\frac{\theta}{4}\right)\right) = \cos\left(\frac{\theta}{2}\right)\\
    &\cos\left(\frac{\theta}{4}\right)\sin\left(\frac{\theta}{4}\right) = \frac{1}{2}\sin\left(\frac{\theta}{2}\right)\\
    &\sin^2\left(\frac{\theta}{4}\right) = \frac{1 - \cos\left(\frac{\theta}{2}\right)}{2} 
\end{align}
as well as properties from the main text.

\subsection*{Overlap Expectation Values}
For completeness we show here to reformulate absolute squares of overlaps as expectation values (see also \cite{lee2018generalized}):
\begin{align}
    S^2_{ij} &= \vert\braket{i}{j}\vert^2\nonumber\\ 
    &= \braket{i}{j}\braket{j}{i} \nonumber\\
    &= \bra{i}U_j\ket{0}\bra{0}U_j^\dagger \ket{i} \nonumber \\
    & = \langle P_0 \rangle_{U_j^\dagger U_i}.
\end{align}
Excited states can be found by applying the variational principle to a projected hamiltonian $QH$ where the $Q =  (1 - \ket{\Psi_i}\bra{\Psi_i} $ projects out already converged states $\ket{\Psi_i}$ prepared by the unitary $U_i$ (note that the unitary includes the reference preparation). 
Expectation values of the projected Hamiltonian can then be written as sum of expectation values of the original Hamiltonian and squares of overlaps of the current circuit $U$
\begin{align}
    \langle QH \rangle_{U} &= \langle H \rangle_{U} - E_i \langle P_0 \rangle_{U_i^\dagger U}
\end{align}

\subsection*{Initialization circuits for CIS states}
The circuits used for the initialization, for the excited states calculation are shown below. For completeness we also show the initialization of a triplet excitation. The circuits are explicit for the Jordan-Wigner representation and would look different for other encodings.
\begin{figure}[h]
    \centering
    \begin{tabular}{cc}
   
    \includegraphics[width=0.24\textwidth]{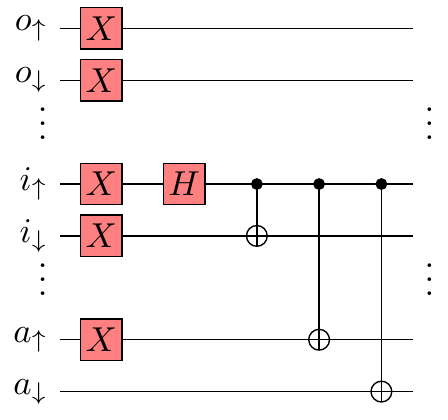}& \includegraphics[width=0.20\textwidth]{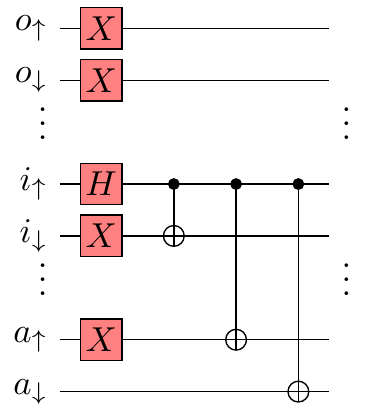}\\
    \end{tabular}
    \caption{Circuits used for the initialization in the excited states calculation. The qubit indices, $o_{\uparrow}$ and $o_{\downarrow}$, represent the occupied orbitals, and the excitation from  $i_{\uparrow}$ and $i_{\downarrow}$ to $a_{\uparrow}$ and $a_{\downarrow}$ represent the two main contributions in the CIS state. 
    Left Panel: The circuit for the Singlet configuration.
    Right Panel: The circuit for the Triplet configuration.}
    \label{fig:cis_init}
\end{figure}

\end{document}